\documentclass[%
 reprint,
 amsmath,amssymb,
 aps,
]{revtex4-2}

\usepackage{graphicx}
\usepackage{dcolumn}
\usepackage{bm}
\usepackage[version=4]{mhchem}
\usepackage{siunitx}
\usepackage{placeins}
\usepackage{caption}

\begin{document}

\title{The third dimension of ferroelectric domain walls}

\author{Erik D. Roede$^1$, Konstantin Shapovalov$^2$, Thomas J. Moran$^3$, Aleksander B. Mosberg$^{4,5}$, Zewu Yan$^{6,7}$, Edith Bourret$^7$, Andres Cano$^8$, Bryan D. Huey$^3$, Antonius T. J. van Helvoort$^4$, and Dennis Meier$^1$}%
 \email{dennis.meier@ntnu.no}
\affiliation{%
$^1$Department of Materials Science and Engineering, Norwegian University of Science and Technology (NTNU), 7491 Trondheim, Norway\\
$^2$Institut de Ciència de Materials de Barcelona (ICMAB-CSIC), Campus UAB, 08193 Bellaterra, Spain\\
$^3$Department of Materials Science and Engineering, University of Connecticut, Storrs, CT 06269, USA.\\
$^4$Department of Physics, Norwegian University of Science and Technology (NTNU), 7491 Trondheim, Norway\\
$^5$SuperSTEM, STFC Daresbury Laboratories, Keckwick Lane, Warrington WA4 4AD, UK\\
$^6$Department of Physics, ETH Zurich, 8093 Zurich, Switzerland \\
$^7$Materials Sciences Division, Lawrence Berkeley National Laboratory, Berkeley, California 94720, USA\\
$^8$Universite Grenoble Alpes, CNRS, Grenoble INP, Institut Néel, Grenoble, France
}%

\date{\today}

\begin{abstract}
Ferroelectric domain walls are quasi-2D systems that show great promise for the development of non-volatile memory, memristor technology and electronic components with ultra-small feature size. Electric fields, for example, can change the domain wall orientation relative to the spontaneous polarization and switch between resistive and conductive states, controlling the electrical current. Being embedded in a 3D material, however, the domain walls are not perfectly flat and can form networks, which leads to complex physical structures. We demonstrate the importance of the nanoscale structure for the emergent transport properties, studying electronic conduction in the 3D network of neutral and charged domain walls in \ce{ErMnO3}. By combining tomographic microscopy techniques and finite element modelling, we clarify the contribution of domain walls within the bulk and show the significance of curvature effects for the local conduction down to the nanoscale. The findings provide insights into the propagation of electrical currents in domain wall networks, reveal additional degrees of freedom for their control, and provide quantitative guidelines for the design of domain wall based technology.

\end{abstract}

\maketitle
Geometric corrugation effects play an important role for the physical properties of 2D materials. In systems such as single-layer graphene and transition-metal dichalcogenides, geometric corrugation phenomena and curvature effects can contribute to the stability and give rise to anisotropic, as well as enhanced mechanical, optical and electronics responses. Fascinating examples are Hall effects in bilayer graphene \cite{Ho2021}  and crystals of artificial atoms forming in \ce{MoS2} \cite{Li2015}, which indicate strong correlations between electrical conduction and deviations from a perfectly flat structure. Recently, ferroelectric domain walls emerged as a completely new type of 2D system with a particularly strong correlation between morphology and electrical responses \cite{Salje2010,Meier2015}. The domain walls exhibit a finite thickness on the order of \SIrange{1}{10}{\angstrom}, because of which they are often referred to as quasi-2D system. Aside from their finite thickness, and analogous to corrugated 2D materials, the walls are not strictly two-dimensional in the sense that they do not develop a perfectly flat structure. Bending and curvature naturally occur to minimize electrostatic stray fields, ensure mechanical compatibility, or due to point defects that lead to domain wall roughening \cite{Yang1999,Tagantsev2010,Paruch2005,Smabraten2020} Importantly, any change in orientation with respect to the electric polarization of the host material directly leads to a modification of the charge state and, hence, the local carrier concentration and conductivity \cite{Eliseev2011,Meier2012} This one-to-one correlation between the domain wall structure and its electronic properties, combined with the spatial mobility of the walls, gives rise to unique functionalities that inspired the emergent field of domain wall nanoelectronics \cite{Catalan2012,Meier2021,Meier2020}. 

In contrast to suspended monolayers, the ferroelectric domain walls are internal 2D systems arising within the bulk, where they separate domains with different orientation of the electric polarization $P$. Thus, their intrinsic electronic transport properties are difficult to access. The conduction of ferroelectric domain walls at surfaces and in near-surface regions has been analyzed by different scanning probe \cite{Seidel2009,Soergel2011,Guy2021,Schultheis2020} and electron microscopy techniques \cite{Jia2007,Pawlik2017,Schaab2014,Becher2015,Nataf2016,Hunnestad2020} and the 3D behavior has been concluded from scans obtained on different surfaces. To resolve the currents that are going through individual ferroelectric domain walls, top and bottom electrodes have been applied to thin films \cite{Seidel2009,Farokhipoor2011,Guyonnet2011}, single-crystals \cite{Sluka2013,Schroeder2012,Meier2012} and FIB-cut lamellas \cite{Mosberg2019}. Such two-terminal measurements have also been combined with domain imaging on the contacted surfaces to correlate the transport behavior with the domain wall orientation relative to $P$, but without resolving the sub-surface structure.

Determining the 3D domain wall structure is crucial to gain quantitative information and understand the complex nanoscale physics as reflected, for example, by the investigations on \ce{Pb(Zr_{0,2}Ti_{0,8})O3} (PZT). PZT thin films with out-of-plane $P$ exhibit enhanced conductance at their nominally neutral \SI{180}{\degree} domain walls, which has been proposed to arise from local head-to-head sections (positively charged) within the material \cite{Jia2007,Guyonnet2011,Tselev2016}. Similarly, the metallic conductance at the nominally neutral domain walls of nano-domains in PZT has been attributed to pronounced curvature effects, leading to strongly charged wall segments below the surface \cite{Maksymovych2011}. Another more recent example is the observation of conductance modulations at nominally neutral domain walls in \ce{ErMnO3} \cite{Schultheis2020}, which have been argued to arise as the walls within the material bend away from the charge-neutral position.

To fill this gap in characterization and correlate the electronic conduction along ferroelectric domain walls to their 3D physical structure, different microscopy-based approaches have been explored. The general strategy is to combine two complementary imaging techniques using one to determine the domain wall orientation and a second to measure the local conductance. For example, domain imaging on opposite surfaces by piezoresponse force microscopy (PFM) \cite{Schroeder2012} and scanning electron microscopy (SEM) \cite{Mosberg2019} has been applied to \ce{LiNbO3} and \ce{ErMnO3}, respectively, to estimate inclination angles $\alpha$ between the domain wall normal $\vec{n}$ and the direction of $\vec{P}$ $(\alpha=\sphericalangle(\vec{P},\vec{n}))$. Based on the data, the average density of domain wall bound charges was calculated $(\rho = 2 |\vec{P}| \cos{\alpha})$  and compared with local transport data gained by conductive atomic force microscopy (cAFM). Such measurements were successful in demonstrating the predicted correlation between emergent domain wall currents and $\rho$ \cite{Eliseev2011,Gureev2011}, but disregard the nanoscale structure of the walls \cite{Gonnissen2016,Zhang2012,Holtz2017}, prohibiting a quantitative analysis at the relevant length scale. Although higher resolution of the domain wall orientation can be achieved by performing transmission electron microscopy \cite{McConville2020} or by correlating PFM and cAFM scans on domain walls in cross-sectional geometry \cite{Zhang2019}, these approaches remain 2D without resolving the wall propagation in the third dimension. This limitation has been overcome by applying optical microscopy, measuring the 3D domain wall structure in \ce{LiNbO3} \cite{Y.2004,Sheng2010,Kaempfe2014,Godau2017} and \ce{LiTaO3} \cite{CherifiHertel2017,Uesu2007} by Cherenkov and nonlinear second harmonic generation, respectively. Although these measurements represent an important breakthrough regarding the 3D characterization of ferroelectric domain walls, their application is restricted to ferroelectrics with specific optical responses. Furthermore, the resolution limit (currently $\approx \SI{1}{\micro\metre}$) has to be pushed to the nanoscale to access the local variations in orientation that determine the charge state and complementary transport measurements remain vital to characterize the electronic domain wall properties.

In this work, we combine focused ion beam (FIB), scanning electron microscopy (SEM) and scanning probe microscopy (SPM) to resolve the 3D domain wall structure in ferroelectric \ce{ErMnO3} and simultaneously record the transport properties with nanoscale spatial precision. The FIB-SEM tomography data correlate the electronic conduction to the local orientation of the domain walls within a single experiment and facilitate realistic calculations that reveal how injected currents spread within the 3D domain wall network. Effects from curvature are studied by high-resolution topographic atomic force microscopy (AFM) and finite-elements calculations, quantifying the additional variations in electronic conduction that arise whenever the domain wall structure deviates from a perfectly flat plane-like geometry. 

\begin{figure*}
\includegraphics[width=\textwidth]{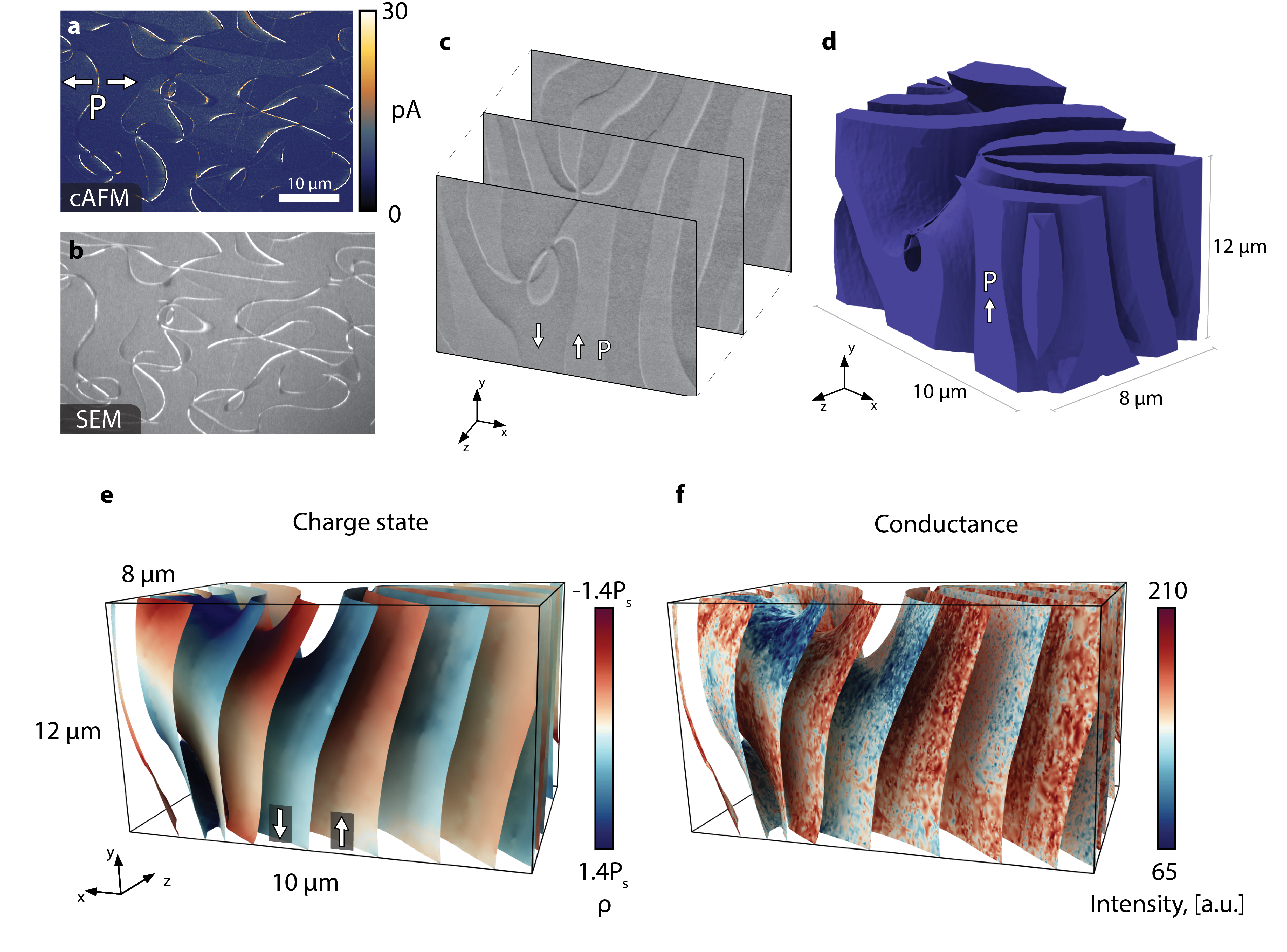}
\caption{\label{fig:1}\textbf{Domain wall charge state and electronic transport properties in 3D.} \textbf{a}, cAFM scan showing conducting and insulating domain walls in \ce{ErMnO3} acquired on a (110)-oriented surface with in-plane polarization ($P\|c$). Arrows indicate the polarization direction of the \SI{180}{\degree} domains. \textbf{b}, SEM image gained at the same position as the cAFM scan in \textbf{a}, showing a one-to-one correlation between the cAFM and SEM contrasts. \textbf{c}, Selected SEM images from the FIB tomography experiment, obtained during the process of cutting \SI{25}{\nano\metre} $z$-slices by FIB. In addition to the domain walls, the SEM data show domain contrast, indicating a non-zero out-of-plane polarization component for the slices shown. \textbf{d}, 3D rendering visualizing the 3D domain structure resolved by FIB tomography. Domains with polarization pointing down ($-P$) are rendered transparent. \textbf{e}, Domain wall orientation and charge state extracted from the volume in \textbf{d}. Colors indicate the domain wall bound charge(blue: positively charged, red: negatively charged) of the domain walls at each point as calculated by projecting the surface normal, $\vec{n}$, of the wall onto the direction of polarization direction. For reference, the density of bound charges at fully charged walls $2P_s$. \textbf{f}, Same domain walls as in \textbf{e}, colored according to the measured SEM intensity; red and blue represent high and low conductivity, respectively.}
\end{figure*}

\section{FIB-SEM tomography for correlated structure and conduction determination in 3D}
For our study we use the uniaxial ferroelectric \ce{ErMnO3} \cite{Yan2015} $(T_C = 1470 K)$ with a saturation polarization $P_S \approx \SI{6}{\micro\coulomb\per\centi\metre\squared}$ oriented along the $c$-axis of the hexagonal unit cell (see Methods for details) \cite{Chae2012,Aken2004}. The material develops a 3D network of meandering domain walls, containing topologically protected vortex structures between which the walls continuously change their charge state as seen in Fig. \ref{fig:1}a \cite{Choi2010,Meier2012}. It is established that wherever the domain walls come to the surface, their electronic transport properties are dominated by the domain wall bound charges \cite{Meier2012}. The latter gives rise to sections that behave insulating (positively charged, head-to-head), bulk-like (neutral, side-by side), and conducting (negatively charged, tail-to-tail) as reflected by the correlated cAFM and SEM data in Fig. \ref{fig:1}a and b \cite{Choi2010,Meier2012,Wu2012}. Although the basic nanoscale physics of the domain walls are well understood, it remains unclear how the injected currents propagate through the 3D domain wall network within the bulk and to what extend the hidden sub-surface structures determine the measured electronic responses. Thus, \ce{ErMnO3} represents an ideal model system for studying the impact of the 3D nanoscale structure and emergent corrugation phenomena on the transport properties at ferroelectric domain walls.

In order to resolve both the 3D domain wall morphology and the local electronic conduction, we combine SEM imaging and FIB tomography, cutting sequential cross-sections into an \ce{ErMnO3} single crystal with out-of-plane $P$. This procedure leads to the image stack presented in Fig. \ref{fig:1}c, where $x$, $y$, and $z$ represent the laboratory coordinate system such that the imaged cross sections are in the $xy$ plane, with polarization oriented in-plane. Consistent with previous SEM surface studies \cite{Roede2021}, domains with opposite polarization orientation are visible as bright ($+P$) and dark ($-P$) regions. In addition, throughout the image series we observe pronounced domain wall contrast, which correlates with their transport properties as reported in ref. \cite{Mosberg2019} and demonstrated in Fig. \ref{fig:1}a and b. For the applied imaging conditions (Methods), the conducting tail-to-tail domain walls are brighter than the ferroelectric domains, whereas the insulating head-to-head domain walls are darker, representing a measure for the local electronic conductance. In total, a volume of $13.8$x$9.2$x$\SI{10.9}{\micro\metre\cubed}$ is imaged with 437 slices and a pitch of \SI{25}{\nano\metre}. The intensity voxel matrix from the tomography data is then reconstructed as explained in Methods, which allows for generating a 3D mesh and selectively visualizing the full 3D domain and domain wall structures as presented in Fig. \ref{fig:1}d to 1f. Figure \ref{fig:1}d presents a 3D view of the domain distribution in \ce{ErMnO3}. The $+P$ domains are colored purple and $-P$ domains are rendered invisible, revealing the curvature of the different domain walls in sub-surface regions and the presence of otherwise hidden vortex–anti-vortex pairs where the domain walls meet as explained elsewhere \cite{Choi2010}.

\begin{figure*}
\includegraphics[width=\textwidth]{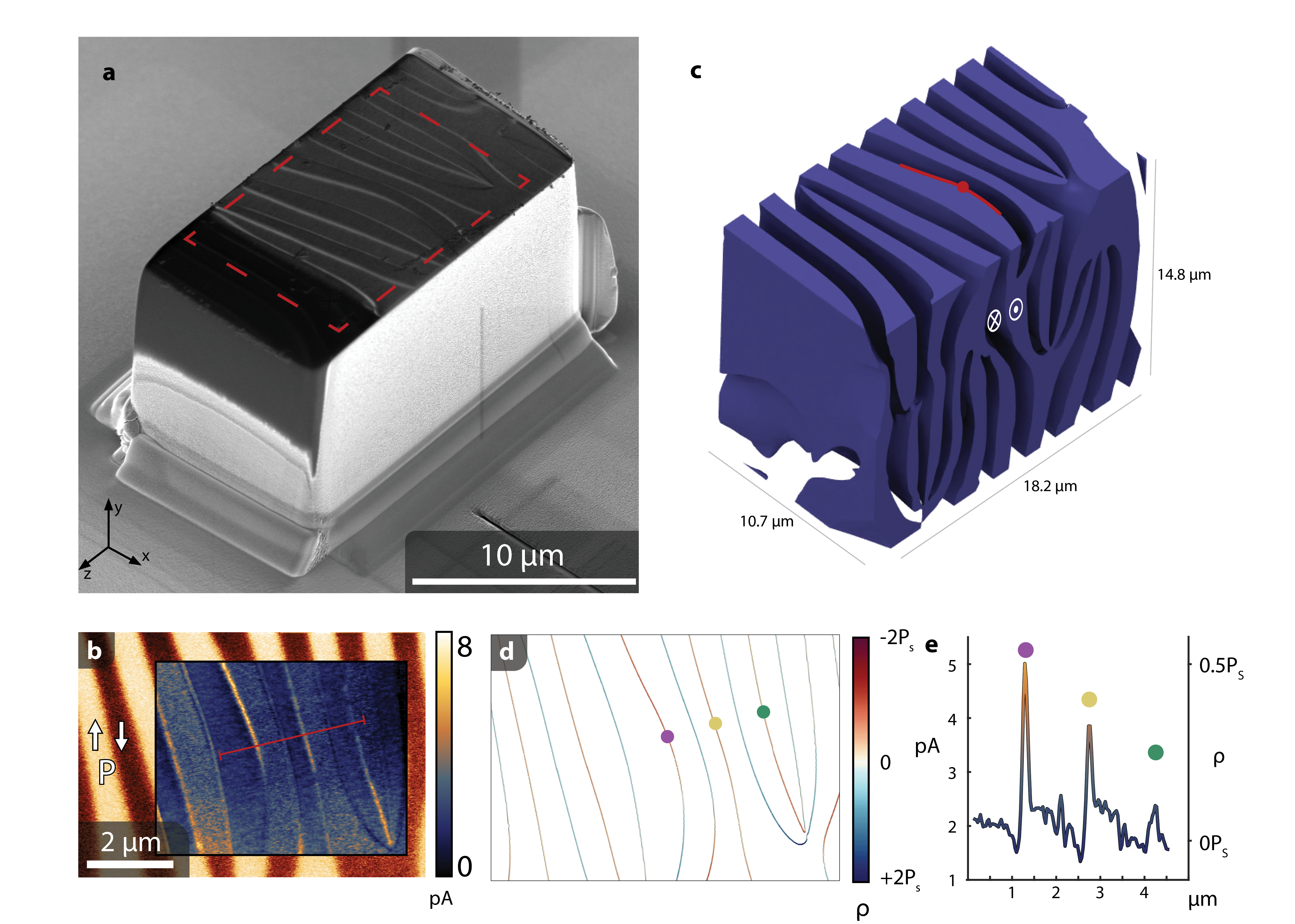}
\caption{\label{fig:2} \textbf{Correlation between domain wall conductance and 3D inclination angle.} \textbf{a}, Sample prepared by FIB liftout, attached to a conductive substrate with top surface polished using \SI{5}{\kilo\volt} \ce{Ga+} ions. \textbf{b}, PFM (in-plane contrast) scan from the region marked by dashed line in \textbf{a}, showing ferroelectric domains with polarization directions marked by white arrows. \textbf{Inset:} cAFM obtained at \SI{10}{\volt} showing conductive domain walls. \textbf{c}, Reconstructed volume from FIB tomography of the sample shown in \textbf{a} with one polarization domain rendered transparent and polarization directions marked by the white symbols (the red line and dot mark the domain wall and point of current injection investigated in the simulations in Fig. \ref{fig:3}). \textbf{d}, Slice from the top of the volume shown in \textbf{c} corresponding to the surface imaged in \textbf{b}, colored according to the 3D charge state (as shown in Fig. \ref{fig:1} e). The colored dots correspond to the wall positions measured in \textbf{b}. \textbf{e}, Line plot showing current cross section from a cAFM scan along the red line in \textbf{b} over three conductive domain walls (averaged over 10 pixels width). The three walls have the same local orientation at the surface, but exhibit different conductance. The colored dots show the real charge state as determined from the 3D geometry in \textbf{c}, confirming that the domain walls in fact have different charge densities.}
\end{figure*}

Based on the 3D data, the charge state at every point on the domain wall network is calculated. For this purpose, the local domain wall normal, $\vec{n}$, is projected onto the direction of $\vec{P}$, which leads to the map of domain wall bound charges in Fig. \ref{fig:1}e (positively charged: blue; negatively charged: red). This tomography-based procedure for evaluating the domain wall charge state is analogous to earlier optical approaches \cite{Kaempfe2014,Godau2017}, but the application of SEM imaging substantially enhances the spatial resolution, providing a much more accurate picture of the 3D morphology and electrostatic potential landscape at the nanoscale.

\newpage
Most importantly, as the SEM intensity is determined by the domain wall conductance (Fig. \ref{fig:1}a, b) \cite{Mosberg2019} the transport properties are mapped simultaneously along with the domain wall orientation. The respective 3D conductance map is presented in Fig. \ref{fig:1}f, with high (red) and low (blue) SEM intensity indicating enhanced and reduced electronic conduction, respectively, compared to the domains. The comparison of the data in Fig. \ref{fig:1}e and \ref{fig:1}f shows a one-to-one correlation between the 3D representation of $\vec{n}$ and the local conductance. This result expands previous transport studies on charged domain walls into the third dimension and highlights the importance of the sub-surface structure which defines the out-of-plane component of $\vec{n}$, co-determining the local conductance.

To gain quantitative insight and understand how the 3D structure of ferroelectric domain walls impacts the response measured in 2D conductance maps, we FIB-cut a micrometer-sized sample out of an \ce{ErMnO3} crystal as displayed in Fig. \ref{fig:2}a ($V = 10.7$x$14.8$x$\SI{18.2}{\micro\metre\cubed}$). In contrast to Fig. \ref{fig:1}, where the probed volume is surrounded by other domains, this geometry ensures that injected currents propagate through a fully characterized domain wall network. The comparison of the data in Fig. \ref{fig:2}a with PFM and cAFM scans from the top surface (Fig. \ref{fig:2}b) shows that the SEM contrast correlates with the domain wall conductance, analogous to the experiments performed on macroscopic samples (Fig. \ref{fig:1}). The corresponding 3D domain structure resolved by FIB-SEM tomography is presented in Fig. \ref{fig:2}c (voxels: $10$x$10$x$\SI{10}{\nano\metre\cubed}$). Our tomography data reveals that several vortex lines with strongly curved domain walls are present within the volume. Furthermore, as the data gives the full 3D domain wall orientation (shown for the surface plane in Fig. \ref{fig:2}d), geometry-related uncertainties in the cAFM data (Fig. \ref{fig:2}b) become quantifiable. Figure \ref{fig:2}e compares the conduction measured at three tail-to-tail domain walls, indicated by the red line in Fig. \ref{fig:2}b. Although the walls seemingly exhibit the same local orientation with respect to $\vec{P}$, substantial variations in domain wall conductance are observed ranging from $25\%$ to $150\%$ higher currents than for the domains. This difference clearly demonstrates the impact of the sub-surface orientation and the 3D inclination angle that determines the actual charge state. The latter is given by the data points in Fig. \ref{fig:2}e, showing the same trend as the local conductance. Note that the domain walls in this example are rather straight. Deviations in the estimation of $\rho$ will be even more pronounced when considering strongly curved domain walls as observed, e.g., close to vortex points in \ce{ErMnO3} (Fig. \ref{fig:2}) \cite{Holtz2017} and nano-domains in PZT \cite{Maksymovych2011} and \ce{BiFeO3} \cite{Vasudevan2012}.

\section{Current injection into 3D domain wall networks}

In the next step, we investigate how injected currents spread within the 3D domain wall network to determine the volume fraction that controls the electronic response observed at the surface. For this purpose, we consider the tail-to-tail domain wall highlighted by the red line in Fig. \ref{fig:2}c and model the spreading of current injected into the structure at the position of the red dot. The conductivity, $\sigma$, is modified at domain walls and vortex lines with respect to the domains, being proportional to the local density of mobile carriers. The latter is approximated as $\sigma = \sigma_\mathrm{bulk} \exp{ ( \nabla \cdot \vec{P} / s )}$, where $s$ defines the maximum conductivity, which arises at the fully charged tail-to-tail domain walls \cite{Meier2012,Wu2012}. Figure 3a presents both the calculated current distribution and the measured charge state of the domain walls in 3D. We find that even for the domain walls with enhanced electronic conduction, a substantial amount of current is flowing through the adjacent domains, consistent with previous analytical solutions for simplified wall geometries \cite{Meier2012}. The relative conductance at domain walls, $G_\mathrm{DW}⁄G_\mathrm{bulk}$ , is dominated by two dimensionless parameters, i.e., the ratio $r / \xi$ between contact radius ($r=\SI{4}{\nano\metre}$ \cite{Abdollahi2019}) and wall width ($\xi = \SI{1}{\nano\metre}$ \cite{Holtz2017}) and the ratio between the intrinsic domain wall conductivity and that of the bulk, $\sigma_\mathrm{DW}^\mathrm{max} / \sigma_\mathrm{bulk}$. As measurements of the intrinsic conductivity at ferroelectric domain walls remain an open and particularly challenging task \cite{Gregg2022}, the latter is treated as a phenomenological fitting parameter in our model. For $\sigma_\mathrm{DW}^\mathrm{max} / \sigma_\mathrm{bulk} = 150$, a realistic conductance of $G_\mathrm{DW}=7.5G_\mathrm{bulk}$ is obtained for a tail-to-tail domain wall oriented perpendicular to the surface, which is in reasonable agreement with local conductance measurements \cite{Meier2012,Schaab2015} and previous simulations performed for a much smaller $r/ \xi$ ratio \cite{Meier2012}. Considering the real domain wall geometry and the injection point marked in Fig. \ref{fig:2}c, we find $G_\mathrm{DW}=1.4G_\mathrm{bulk}$.

 \begin{figure}
\includegraphics[width=0.48\textwidth]{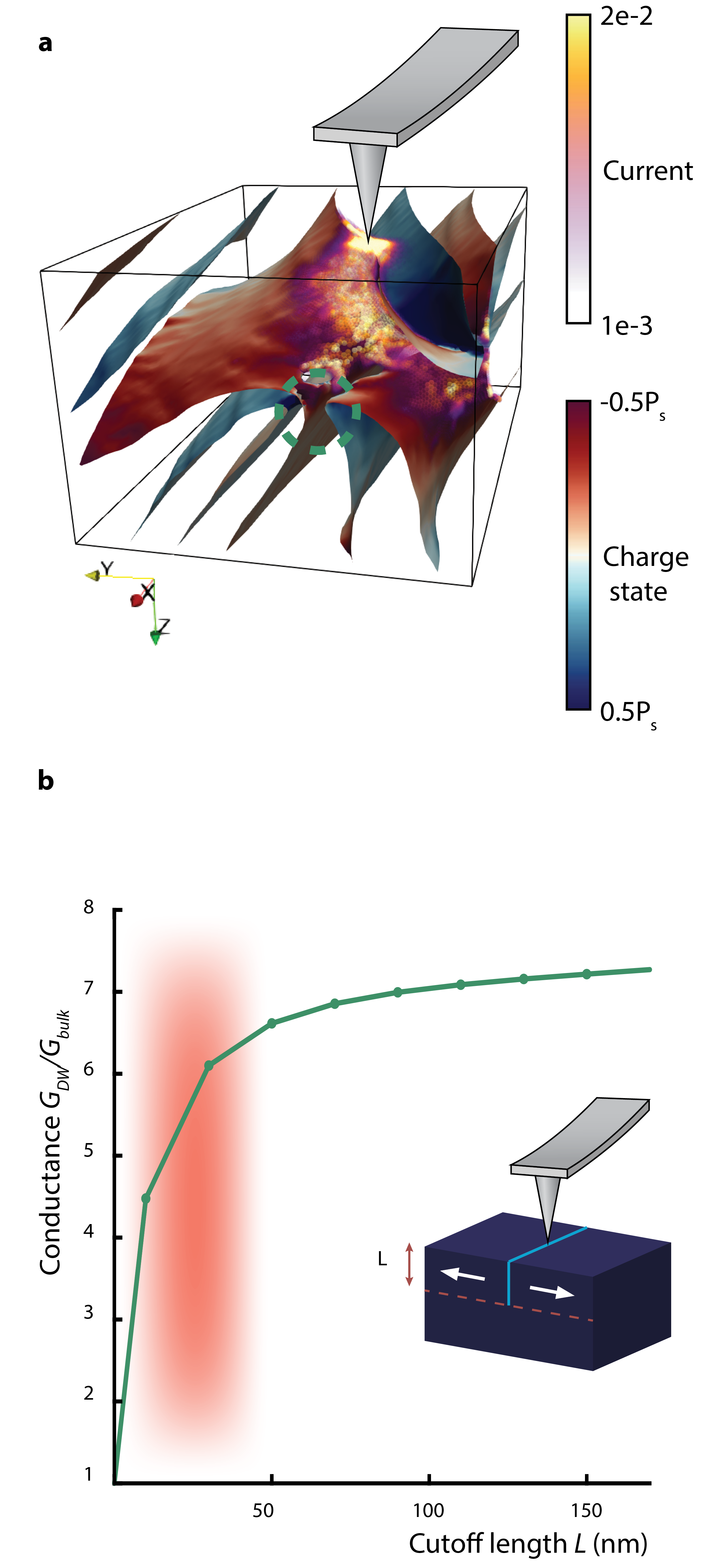}
\caption{\label{fig:3} \textbf{Current spreading in a 3D domain wall network.} \textbf{a}, Calculated distribution of electrical current injected into the domain wall network in Fig. 2c at the position marked by the red dot. The current spreads over different walls with the vortex (green dashed circle) acting as current divider. \textbf{b}, Calculated normalized conductance as a function of the cut-off length, $L$. The cut-off length is defined as illustrated in the inset, representing a measure for the domain wall segment that dominates the conduction properties. The data shows that 90\% of the measured conductance originate from a depth $\leq \SI{50}{\nano\metre}$}
\end{figure}

Most importantly, Fig. \ref{fig:3}a shows that the injected current flows through more than one wall, representing a superposition of different walls with continuously varying charge states. In hexagonal manganites, additional complication potentially arises from the presence of topologically protected vortices as seen in the center of Fig. \ref{fig:3}a (highlighted by the dashed circle), which can act as current dividers within the domain wall network. Because of the complexity of the current distribution in Fig. \ref{fig:3}a, it is clear that local transport measurements gained at the surface yield only a very rough estimate. This is a general problem for the research on domain walls, hampering the development of more precise models that allow for disentangling the different emergent conduction contributions \cite{Gregg2022,Schultheis2021}. However, Fig. \ref{fig:3}a also reveals that the current density quickly drops off with increasing distance from the surface. For the ideal case of a completely homogeneous material, the electric potential can be estimated to drop by 75\% over a distance $d\approx\SI{8}{\nano\metre}$ for a tip-surface contact radius of $r\approx\SI{4}{\nano\metre}$ \cite{LJ1957, holm1958}. 

To gain a realistic estimate for the domain walls in \ce{ErMnO3}, we consider a tail-to-tail domain wall modelled as a conducting 2D channel with varying penetration depths as illustrated in the inset to Fig. \ref{fig:3}b. The calculated conductance $G_\mathrm{DW}$ is then a function of the cut-off length, $L$, normalized to the conductance of the bulk $G_\mathrm{bulk}$. The result demonstrates that for a fully charged tail-to-tail wall, 90\% of the conductance signal originates from a depth $\leq \SI{50}{\nano\metre}$. As the conductance plateaus at $L_c \approx \SI{50}{\nano\metre}$ the domain wall geometry and emergent vortex lines further away from the surface are practically irrelevant and the sample essentially behaves like an ohmic resistor. We note that $L_c$ quickly decreases as domain walls tilt away from the fully charged tail-to-tail states; for a partially charged tail-to-tail domain wall inclined by \SI{45}{\degree} with respect to $P$ $\sigma_\mathrm{DW}/\sigma_\mathrm{bulk} = 12$  we find $L_c \approx \SI{15}{\nano\metre}$ (see Supplementary Figure \ref{fig:S2}). Vice versa, in systems with higher domain wall conductivities – such as \ce{LiNbO3}, where the conductivity is enhanced by up to 13 orders of magnitude \cite{Werner2017}– substantially larger $L_c$ values can be expected.

In summary, Fig. \ref{fig:3} demonstrates that the existence of a domain-wall-specific cut-off length, $L_c$, which is determined by the relative conductance $G_\mathrm{DW}⁄G_\mathrm{bulk}$ . Within this cut-off length, the domain wall charge state and connections with other domain walls are critical, controlling the current path and hence the measured domain wall conductivity. This observation highlights the importance of the near-surface domain wall nanostructure beyond just the 3D inclination angle, including corrugation phenomena and curvature effects promoted by, e.g., the surface termination, local electrostatics, and strain. Furthermore, it suggests an additional and even stronger mechanism for controlling the transport properties at domain walls, leveraging permanent or transient changes in curvature rather than by varying the domain wall position or overall tilt angle.

\section{Relation between domain wall curvature and electronic transport}
\begin{figure*}
\includegraphics[width=\textwidth]{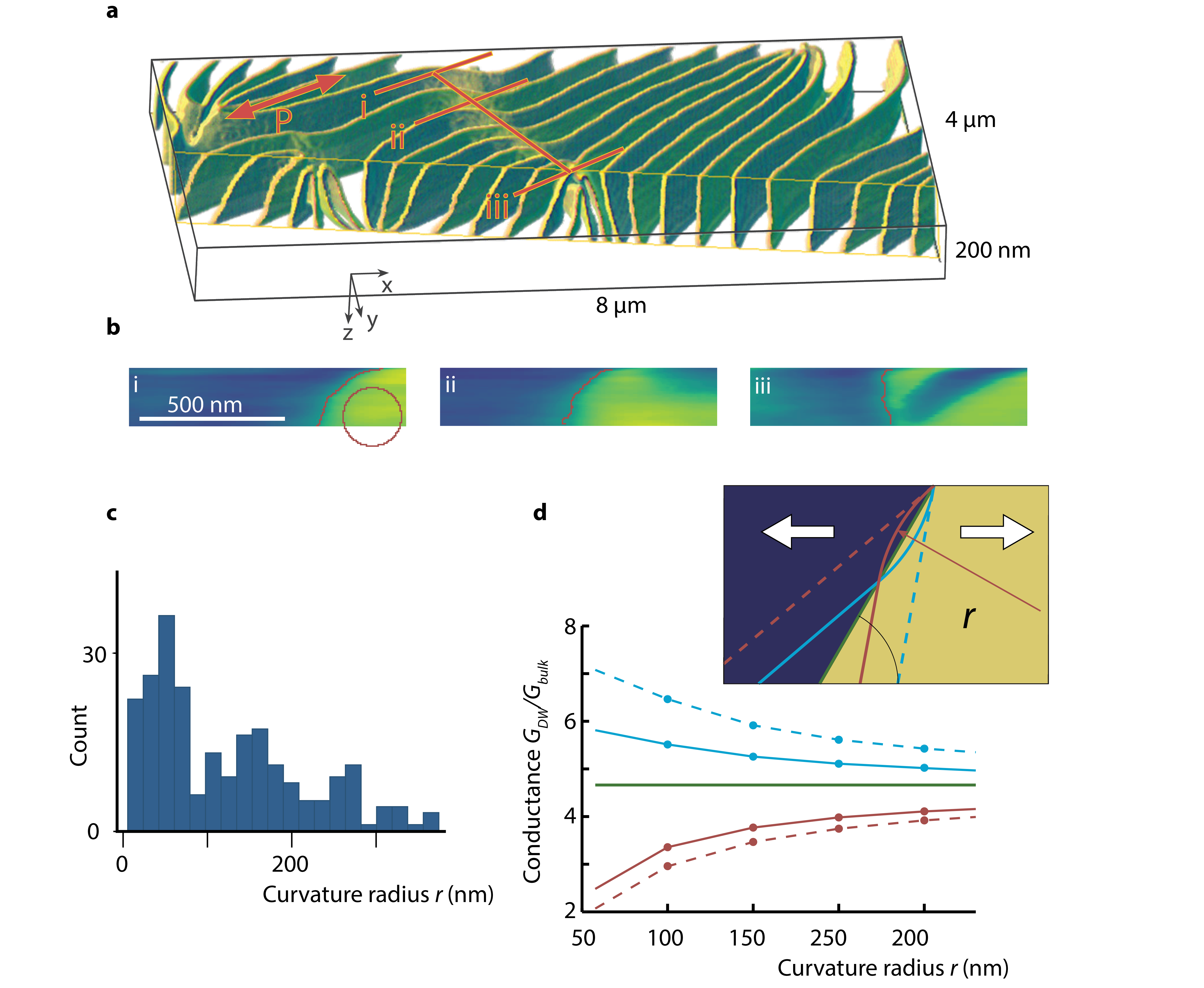}
\caption{\label{fig:4}\textbf{Relation between domain wall curvature and conductance.} \textbf{a}, Tomographic PFM data revealing the 3D structure of ferroelectric domain walls in the near-surface region. Red lines indicate the three positions corresponding to the cross-sectional images shown in \textbf{b}. \textbf{b}, Cross-sectional images revealing the local curvature of three representative domain walls. A circle of radius 100 nm is shown in (i) to illustrate the respective curvature for reference. \textbf{c}, Histogram of domain wall curvature radius. The curvature radius is calculated for each pixel at the walls marked in \textbf{b}, and the counts of resulting values are plotted in the histogram, showing that emergent radii are on the \SI{100}{\nano\metre} length scale. \textbf{d}, Relation between domain wall curvature and conductance. The simulated domain wall geometries are shown in the inset: a planar tail-to-tail domain wall with \SI{60}{\degree} inclination angle relative to the surface (green), convex/concave curved walls (red and blue) with the same average angle near the surface as the green wall, and straight walls tangent to the curved walls at the surface (dashed red and blue).}
\end{figure*}

To explore and quantify the nanoscale curvature-driven conduction contributions that arise in \ce{ErMnO3}, we resolve the near-surface domain wall nanostructure using a PFM version of high-resolution tomographic AFM \cite{Song2021} and calculate resulting variations in conductivity. Tomographic PFM facilitates ultra-high volumetric resolution of domains \cite{Song2021a,Steffes2019}, in this case implementing voxels of $15.6$x$15.6$x$\SI{5}{\nano\metre\cubed}$, making it an ideal tool for determining near-surface domain wall tilts and curvature. Figure \ref{fig:4}a presents a 3D tomogram of all domain walls within an $8$x$4$x$\SI{0.2}{\micro\metre\cubed}$ block of excavated material from the same batch as investigated in Fig. \ref{fig:1} to \ref{fig:3}. Based on the data, substantial and heterogeneous nanoscale curvature is confirmed not just at the surface but also into the depth. 

Figure 4b demonstrates three representative cross-sections extracted from the original piezoresponse tomogram. Overlays of the domain wall location into the depth are also shown, along with a $\SI{100}{\nano\metre}$ radius of curvature circle to guide the eye. Figure 4c presents a histogram of such curvatures calculated for a total of 231 points from 3 cross sections. There are many regions which locally exhibit curvatures less than $\SI{100}{\nano\metre}$, a broader peak around $\SI{150}{\nano\metre}$, and visually there are many sections without extensive curvature along the $z$ axis. If such various domain wall structures can be engineered or even temporally controlled, then their influence on the conductance could be profound according to Fig. \ref{fig:4}d. This plot reveals the range of conductances achievable for tail-to-tail domain walls that are either tilted or curved as defined in the inset. Specifically, variations in the electronic transport are evaluated (see Supplementary Note 1) assuming a current is injected into three differently shaped tail-to-tail walls (convex, concave and tilted straight wall) which symmetrically traverse the same point at a depth of $\SI{50}{\nano\metre}$. Beyond this point, all walls are assumed to continue as straight walls. This setting allows us to quantify the impact of the curvature for domain walls that nominally exhibit the same average inclination angle, here \SI{60}{\degree}, focusing on the near-surface region which dominates the measured conductance ($> 90\%$ for $L_c = \SI{50}{\nano\metre}$, Fig. \ref{fig:3}b). The simulation reveals two important effects: Firstly, the curvature of the wall plays a key role in its conductance, leading to substantially different values than calculated from the 3D inclination angle at the surface (see Fig. \ref{fig:2}e) or the inclination angle averaged over the investigated near-surface layer. For example, for a moderately curved concave tail-to-tail wall ($r=\SI{100}{\nano\metre}$), the extracted conductance is $15\%$ lower than for a straight wall with the same 3D inclination angle at the surface (\SI{77}{\degree} for the dashed blue line) and $18\%$ higher than the green straight wall that goes through the same point at $\SI{50}{\nano\metre}$ depth (solid green line). Secondly, the direction of curvature has a pronounced influence on the domain wall conductance. For curvature radii similar to those resolved by the domain wall tomogram of Fig. \ref{fig:4}a-b, our model indicates a $60\%$ higher conductance at concave tail-to-tail walls compared to convex ones ($r=\SI{100}{\nano\metre}$). These results provide quantitative insight into curvature-driven conduction phenomena, reveal corrugation as an important parameter for controlling the electronic response at ferroelectric domain walls, and suggest a new approach for the design and optimization of domain wall devices. 

\section{Conclusions}

The 3D investigations in this work demonstrate the as-yet-unused non-planar nature of ferroelectric domain walls as important parameter for controlling the local transport behavior. By leveraging either permanent or temporary variations in curvature, the domain wall resistance can readily be controlled without the need to completely change the domain wall position or its overall tilt angle. This possibility gives a new dimension to domain wall engineering, enabling the design of two-terminal devices, where continuous variations in conductance could be achieved via static or dynamical curvature effects. Our investigations give detailed information concerning emergent curvature-driven transport phenomena and outline basic criteria for device design, including a practical upper limit for electrode-electrode distances as defined by the cut-off length $L_c$. The opportunity to generate continuous conductivity changes via curvature effects allows for multi-level resistance control and complex electronic responses based on individual domain walls. With this, we expand the field of domain wall electronics into the realm of nanoscale synaptic devices and unconventional computing, utilizing single ferroelectric domain walls as ultra-small units that facilitate the desired discrete and tunable functionalities.

\subsection{Acknowledgements}
 The authors acknowledge NTNU for support through the Enabling Nanotechnologies: Nano program. The Research Council of Norway is acknowledged for the support to the Norwegian Micro- and Nano-Fabrication Facility, NorFab, project number 245963/F50. ZY and EB were supported by the U.S. Department of Energy, Office of Science, Basic Energy Sciences, Materials Sciences and Engineering Division under Contract No. DE-AC02-05-CH11231 within the Quantum Materials program-KC2202. DM acknowledges funding from the European Research Council (ERC) under the European Union’s Horizon 2020 research and innovation program (Grant Agreement No. 863691) and further thanks NTNU for support through the Onsager Fellowship Program and NTNU Stjerneprogrammet. TJM and BDH are grateful for support from NSF DMR:MRI:1726862 to develop T-AFM. KS acknowledges the support of the European Research Council under the European Union’s Horizon 2020 research and innovation program (Grant Agreement No. 724529) and Ministerio de Economia, Industria y Competitividad through Grant Nos. MAT2016-77100-C2-2-P and SEV-2015-0496.

\subsection{Author contributions}
 EDR recorded the PFM/cAFM data and performed the FIB tomography measurements together with ABM under the supervision of DM and ATJvH. KS performed the 3D finite-elements modelling of the current flow supervised by MS, with support from AC. TJM and BDH conducted and analyzed tomographic AFM. ZY and EB provided the materials. DM coordinated the project and wrote the manuscript together with EDR. All authors discussed the results and contributed to the final version of the manuscript.

\section{Methods}

\subsection{Samples} 
High-quality single crystals of \ce{ErMnO3} were prepared by the pressurized floating-zone method \cite{Yan2015}, oriented by Laue diffraction and cut to disks of about \SI{1}{\milli\metre} thickness. The surfaces were chemomechanically polished using a Logitech PM5 polishing machine with silica slurry. The polished surface had a root-mean-square roughness of $\approx$ \SI{1}{\nano\metre}.

\subsection{FIB-SEM tomography and liftout} 
A Thermo Scientific Helios G4 UX DualBeam FIB-SEM was used for imaging, tomography and nanostructuring. Scanning electron microscopy (SEM) images were taken at \SI{2}{\kilo\volt} acceleration voltage at \SI{0.1}{\nano\ampere} beam current, using an in-lens detector with an immersion objective lens. For FIB liftout, a pentagonal prism was extracted in a plan view configuration using an EasyLift EX NanoManipulator and attached to a TEM half grid. The sample was then trimmed to leave a rectangular prism. The bottom face was polished using \SI{5}{\kilo\volt} acceleration voltage before the sample was lifted off the TEM grid and \ce{Pt}-welded to a gold-coated \ce{Si} substrate and the top face polished at \SI{5}{\kilo\volt} . Before tomography, areas of interest were coated with \SI{2}{\micro\metre} ion beam deposited platinum. The Slice and View 4 software was used for automated serial sectioning. For milling, \SI{30}{\kilo\volt}  ion beam acceleration voltage and $90$ to \SI{260}{\pico\ampere} beam current was used. 

\subsection{Tomography image processing} 
FIB based tomography image stacks were aligned using the software ImageJ \cite{Rueden2017,Schindelin2012} by applying a template matching algorithm to superimpose known fixed points. To account for charging artifacts that may occur during SEM imaging, a median line alignment was implemented, adjusting the brightness of each line to equalize the median of each line. To counter the effect of apparent changes in illumination during the tomography procedure, a pseudo flat-field background subtraction was performed by creating and subtracting an artificial background image for each slice. After semiautomated segmentation into different polarization domains, the extracted voxel matrices were processed in ParaView \cite{Ayachit2015}. A contouring filter was used to extract the domain walls as a 3D mesh, which was smoothed to reduce staircasing artifacts.

\subsection{Scanning probe microscopy} 
Conductive atomic force microscopy (cAFM) and piezoresponse force microscopy (PFM) measurements were performed on an Oxford Instruments Cypher ES AFM. For cAFM, TipsNano DCP10 probes were used with a \SI{10}{\volt} bias applied to the back of the sample. For PFM, RMN 25PT300B probes were used in lateral dual AC resonance tracking mode around the contact resonance of \SI{229.45}{\kilo\hertz}, with \SI{10}{\volt} peak-to-peak AC voltage applied. The polarization directions were calibrated on a PPLNC (periodically poled \ce{LiNbO3}). 

\subsection{AFM tomography} 
Consecutive lateral PFM images were acquired with an Oxford Instruments Cypher VRS AFM, using doped diamond AFM probes (Appnano DD-ACTA) and a ZurichInstruments MFLI lock-in-ampifier. Tip biases on the order of \SI{2}{\volt}, at torsional contact resonant frequencies of $\approx 1 - \SI{2}{\mega\hertz}$, provided PFM amplitude and phase images as typified by the six representative phase images out of 41 total in Supplementary Fig. \ref{fig:S1}a. From these, the piezoresponse is calculated pixel by pixel throughout the entire volume (Supplementary Fig. \ref{fig:S1}b, implemented with Matlab using custom but straightforward code which calculates amplitude*sin(phase). To better visualize the domain curvature, all positive (bright) domains are removed in (Supplementary Fig. \ref{fig:S1}c) and the tomogram is depicted from 3 slightly different perspectives, based on $\SI{-20}{\degree}$, $\SI{0}{\degree}$, and $\SI{20}{\degree}$ rotations around the $y$ axis (as indicated) and with oblique illumination and shadowing (according to a light source shown at lower right). As with the FIB based tomography, $xy$ position registry and then 3D domain wall identification are both performed using ImageJ. Slicing across multiple domain walls particularly reveals extensive local curvature of the domain patterns throughout $x$, $y$, and $z$, confirming the profoundly 3-dimensional nanoscale nature of the domain structure which is otherwise hidden beneath the surface. These results are typical of more than 10 similarly acquired volumes from this same batch of samples as utilized throughout this work.

\subsection{Finite element modelling} 
All simulations of the current flow in presence of domain walls are done in 3D using the finite-elements method implemented in COMSOL Multiphysics – see details in the Supplementary Note 1.

\bibliography{cubeBib}

\begin{thebibliography}{62}%
\makeatletter
\providecommand \@ifxundefined [1]{%
 \@ifx{#1\undefined}
}%
\providecommand \@ifnum [1]{%
 \ifnum #1\expandafter \@firstoftwo
 \else \expandafter \@secondoftwo
 \fi
}%
\providecommand \@ifx [1]{%
 \ifx #1\expandafter \@firstoftwo
 \else \expandafter \@secondoftwo
 \fi
}%
\providecommand \natexlab [1]{#1}%
\providecommand \enquote  [1]{``#1''}%
\providecommand \bibnamefont  [1]{#1}%
\providecommand \bibfnamefont [1]{#1}%
\providecommand \citenamefont [1]{#1}%
\providecommand \href@noop [0]{\@secondoftwo}%
\providecommand \href [0]{\begingroup \@sanitize@url \@href}%
\providecommand \@href[1]{\@@startlink{#1}\@@href}%
\providecommand \@@href[1]{\endgroup#1\@@endlink}%
\providecommand \@sanitize@url [0]{\catcode `\\12\catcode `\$12\catcode
  `\&12\catcode `\#12\catcode `\^12\catcode `\_12\catcode `\%12\relax}%
\providecommand \@@startlink[1]{}%
\providecommand \@@endlink[0]{}%
\providecommand \url  [0]{\begingroup\@sanitize@url \@url }%
\providecommand \@url [1]{\endgroup\@href {#1}{\urlprefix }}%
\providecommand \urlprefix  [0]{URL }%
\providecommand \Eprint [0]{\href }%
\providecommand \doibase [0]{https://doi.org/}%
\providecommand \selectlanguage [0]{\@gobble}%
\providecommand \bibinfo  [0]{\@secondoftwo}%
\providecommand \bibfield  [0]{\@secondoftwo}%
\providecommand \translation [1]{[#1]}%
\providecommand \BibitemOpen [0]{}%
\providecommand \bibitemStop [0]{}%
\providecommand \bibitemNoStop [0]{.\EOS\space}%
\providecommand \EOS [0]{\spacefactor3000\relax}%
\providecommand \BibitemShut  [1]{\csname bibitem#1\endcsname}%
\let\auto@bib@innerbib\@empty
\bibitem [{\citenamefont {Ho}\ \emph {et~al.}(2021)\citenamefont {Ho},
  \citenamefont {Chang}, \citenamefont {Hsieh}, \citenamefont {Lo},
  \citenamefont {Huang}, \citenamefont {Vu}, \citenamefont {Ortix},\ and\
  \citenamefont {Chen}}]{Ho2021}%
  \BibitemOpen
  \bibfield  {author} {\bibinfo {author} {\bibfnamefont {S.-C.}\ \bibnamefont
  {Ho}}, \bibinfo {author} {\bibfnamefont {C.-H.}\ \bibnamefont {Chang}},
  \bibinfo {author} {\bibfnamefont {Y.-C.}\ \bibnamefont {Hsieh}}, \bibinfo
  {author} {\bibfnamefont {S.-T.}\ \bibnamefont {Lo}}, \bibinfo {author}
  {\bibfnamefont {B.}~\bibnamefont {Huang}}, \bibinfo {author} {\bibfnamefont
  {T.-H.-Y.}\ \bibnamefont {Vu}}, \bibinfo {author} {\bibfnamefont
  {C.}~\bibnamefont {Ortix}},\ and\ \bibinfo {author} {\bibfnamefont {T.-M.}\
  \bibnamefont {Chen}},\ }\bibfield  {title} {\bibinfo {title} {Hall effects in
  artificially corrugated bilayer graphene without breaking time-reversal
  symmetry},\ }\href {https://doi.org/10.1038/s41928-021-00537-5} {\bibfield
  {journal} {\bibinfo  {journal} {Nature Electronics}\ }\textbf {\bibinfo
  {volume} {4}},\ \bibinfo {pages} {116} (\bibinfo {year} {2021})}\BibitemShut
  {NoStop}%
\bibitem [{\citenamefont {Li}\ \emph {et~al.}(2015)\citenamefont {Li},
  \citenamefont {Contryman}, \citenamefont {Qian}, \citenamefont {Ardakani},
  \citenamefont {Gong}, \citenamefont {Wang}, \citenamefont {Weisse},
  \citenamefont {Lee}, \citenamefont {Zhao}, \citenamefont {Ajayan},
  \citenamefont {Li}, \citenamefont {Manoharan},\ and\ \citenamefont
  {Zheng}}]{Li2015}%
  \BibitemOpen
  \bibfield  {author} {\bibinfo {author} {\bibfnamefont {H.}~\bibnamefont
  {Li}}, \bibinfo {author} {\bibfnamefont {A.~W.}\ \bibnamefont {Contryman}},
  \bibinfo {author} {\bibfnamefont {X.}~\bibnamefont {Qian}}, \bibinfo {author}
  {\bibfnamefont {S.~M.}\ \bibnamefont {Ardakani}}, \bibinfo {author}
  {\bibfnamefont {Y.}~\bibnamefont {Gong}}, \bibinfo {author} {\bibfnamefont
  {X.}~\bibnamefont {Wang}}, \bibinfo {author} {\bibfnamefont {J.~M.}\
  \bibnamefont {Weisse}}, \bibinfo {author} {\bibfnamefont {C.~H.}\
  \bibnamefont {Lee}}, \bibinfo {author} {\bibfnamefont {J.}~\bibnamefont
  {Zhao}}, \bibinfo {author} {\bibfnamefont {P.~M.}\ \bibnamefont {Ajayan}},
  \bibinfo {author} {\bibfnamefont {J.}~\bibnamefont {Li}}, \bibinfo {author}
  {\bibfnamefont {H.~C.}\ \bibnamefont {Manoharan}},\ and\ \bibinfo {author}
  {\bibfnamefont {X.}~\bibnamefont {Zheng}},\ }\bibfield  {title} {\bibinfo
  {title} {Optoelectronic crystal of artificial atoms in strain-textured
  molybdenum disulphide},\ }\bibfield  {journal} {\bibinfo  {journal} {Nature
  Communications}\ }\textbf {\bibinfo {volume} {6}},\ \href
  {https://doi.org/10.1038/ncomms8381} {10.1038/ncomms8381} (\bibinfo {year}
  {2015})\BibitemShut {NoStop}%
\bibitem [{\citenamefont {Salje}(2010)}]{Salje2010}%
  \BibitemOpen
  \bibfield  {author} {\bibinfo {author} {\bibfnamefont {E.~K.~H.}\
  \bibnamefont {Salje}},\ }\bibfield  {title} {\bibinfo {title} {Multiferroic
  domain boundaries as active memory devices: Trajectories towards domain
  boundary engineering},\ }\href {https://doi.org/10.1002/cphc.200900943}
  {\bibfield  {journal} {\bibinfo  {journal} {{ChemPhysChem}}\ }\textbf
  {\bibinfo {volume} {11}},\ \bibinfo {pages} {940} (\bibinfo {year}
  {2010})}\BibitemShut {NoStop}%
\bibitem [{\citenamefont {Meier}(2015)}]{Meier2015}%
  \BibitemOpen
  \bibfield  {author} {\bibinfo {author} {\bibfnamefont {D.}~\bibnamefont
  {Meier}},\ }\bibfield  {title} {\bibinfo {title} {Functional domain walls in
  multiferroics},\ }\href {https://doi.org/10.1088/0953-8984/27/46/463003}
  {\bibfield  {journal} {\bibinfo  {journal} {Journal of Physics: Condensed
  Matter}\ }\textbf {\bibinfo {volume} {27}},\ \bibinfo {pages} {463003}
  (\bibinfo {year} {2015})}\BibitemShut {NoStop}%
\bibitem [{\citenamefont {Yang}\ \emph {et~al.}(1999)\citenamefont {Yang},
  \citenamefont {Gopalan}, \citenamefont {Swart},\ and\ \citenamefont
  {Mohideen}}]{Yang1999}%
  \BibitemOpen
  \bibfield  {author} {\bibinfo {author} {\bibfnamefont {T.~J.}\ \bibnamefont
  {Yang}}, \bibinfo {author} {\bibfnamefont {V.}~\bibnamefont {Gopalan}},
  \bibinfo {author} {\bibfnamefont {P.~J.}\ \bibnamefont {Swart}},\ and\
  \bibinfo {author} {\bibfnamefont {U.}~\bibnamefont {Mohideen}},\ }\bibfield
  {title} {\bibinfo {title} {Direct observation of pinning and bowing of a
  single ferroelectric domain wall},\ }\href
  {https://doi.org/10.1103/physrevlett.82.4106} {\bibfield  {journal} {\bibinfo
   {journal} {Physical Review Letters}\ }\textbf {\bibinfo {volume} {82}},\
  \bibinfo {pages} {4106} (\bibinfo {year} {1999})}\BibitemShut {NoStop}%
\bibitem [{\citenamefont {Tagantsev}\ \emph {et~al.}()\citenamefont
  {Tagantsev}, \citenamefont {Cross},\ and\ \citenamefont
  {Fousek}}]{Tagantsev2010}%
  \BibitemOpen
  \bibfield  {author} {\bibinfo {author} {\bibfnamefont {A.~K.}\ \bibnamefont
  {Tagantsev}}, \bibinfo {author} {\bibfnamefont {L.~E.}\ \bibnamefont
  {Cross}},\ and\ \bibinfo {author} {\bibfnamefont {J.}~\bibnamefont
  {Fousek}},\ }\href {https://doi.org/10.1007/978-1-4419-1417-0} {\emph
  {\bibinfo {title} {Domains in Ferroic Crystals and Thin Films}}}\ (\bibinfo
  {publisher} {Springer New York})\BibitemShut {NoStop}%
\bibitem [{\citenamefont {Paruch}\ \emph {et~al.}(2005)\citenamefont {Paruch},
  \citenamefont {Giamarchi},\ and\ \citenamefont {Triscone}}]{Paruch2005}%
  \BibitemOpen
  \bibfield  {author} {\bibinfo {author} {\bibfnamefont {P.}~\bibnamefont
  {Paruch}}, \bibinfo {author} {\bibfnamefont {T.}~\bibnamefont {Giamarchi}},\
  and\ \bibinfo {author} {\bibfnamefont {J.-M.}\ \bibnamefont {Triscone}},\
  }\bibfield  {title} {\bibinfo {title} {Domain wall roughness in epitaxial
  ferroelectric \ce{PbZr_{0.2}Ti_{0.8}O3} thin films},\ }\bibfield  {journal}
  {\bibinfo  {journal} {Physical Review Letters}\ }\textbf {\bibinfo {volume}
  {94}},\ \href {https://doi.org/10.1103/physrevlett.94.197601}
  {10.1103/physrevlett.94.197601} (\bibinfo {year} {2005})\BibitemShut
  {NoStop}%
\bibitem [{\citenamefont {Sm{\aa}br{\aa}ten}\ \emph {et~al.}(2020)\citenamefont
  {Sm{\aa}br{\aa}ten}, \citenamefont {Holstad}, \citenamefont {Evans},
  \citenamefont {Yan}, \citenamefont {Bourret}, \citenamefont {Meier},\ and\
  \citenamefont {Selbach}}]{Smabraten2020}%
  \BibitemOpen
  \bibfield  {author} {\bibinfo {author} {\bibfnamefont {D.~R.}\ \bibnamefont
  {Sm{\aa}br{\aa}ten}}, \bibinfo {author} {\bibfnamefont {T.~S.}\ \bibnamefont
  {Holstad}}, \bibinfo {author} {\bibfnamefont {D.~M.}\ \bibnamefont {Evans}},
  \bibinfo {author} {\bibfnamefont {Z.}~\bibnamefont {Yan}}, \bibinfo {author}
  {\bibfnamefont {E.}~\bibnamefont {Bourret}}, \bibinfo {author} {\bibfnamefont
  {D.}~\bibnamefont {Meier}},\ and\ \bibinfo {author} {\bibfnamefont {S.~M.}\
  \bibnamefont {Selbach}},\ }\bibfield  {title} {\bibinfo {title} {Domain wall
  mobility and roughening in doped ferroelectric hexagonal manganites},\
  }\bibfield  {journal} {\bibinfo  {journal} {Physical Review Research}\
  }\textbf {\bibinfo {volume} {2}},\ \href
  {https://doi.org/10.1103/physrevresearch.2.033159}
  {10.1103/physrevresearch.2.033159} (\bibinfo {year} {2020})\BibitemShut
  {NoStop}%
\bibitem [{\citenamefont {Eliseev}\ \emph {et~al.}(2011)\citenamefont
  {Eliseev}, \citenamefont {Morozovska}, \citenamefont {Svechnikov},
  \citenamefont {Gopalan},\ and\ \citenamefont {Shur}}]{Eliseev2011}%
  \BibitemOpen
  \bibfield  {author} {\bibinfo {author} {\bibfnamefont {E.~A.}\ \bibnamefont
  {Eliseev}}, \bibinfo {author} {\bibfnamefont {A.~N.}\ \bibnamefont
  {Morozovska}}, \bibinfo {author} {\bibfnamefont {G.~S.}\ \bibnamefont
  {Svechnikov}}, \bibinfo {author} {\bibfnamefont {V.}~\bibnamefont
  {Gopalan}},\ and\ \bibinfo {author} {\bibfnamefont {V.~Y.}\ \bibnamefont
  {Shur}},\ }\bibfield  {title} {\bibinfo {title} {Static conductivity of
  charged domain walls in uniaxial ferroelectric semiconductors},\ }\bibfield
  {journal} {\bibinfo  {journal} {Physical Review B}\ }\textbf {\bibinfo
  {volume} {83}},\ \href {https://doi.org/10.1103/physrevb.83.235313}
  {10.1103/physrevb.83.235313} (\bibinfo {year} {2011})\BibitemShut {NoStop}%
\bibitem [{\citenamefont {Meier}\ \emph {et~al.}(2012)\citenamefont {Meier},
  \citenamefont {Seidel}, \citenamefont {Cano}, \citenamefont {Delaney},
  \citenamefont {Kumagai}, \citenamefont {Mostovoy}, \citenamefont {Spaldin},
  \citenamefont {Ramesh},\ and\ \citenamefont {Fiebig}}]{Meier2012}%
  \BibitemOpen
  \bibfield  {author} {\bibinfo {author} {\bibfnamefont {D.}~\bibnamefont
  {Meier}}, \bibinfo {author} {\bibfnamefont {J.}~\bibnamefont {Seidel}},
  \bibinfo {author} {\bibfnamefont {A.}~\bibnamefont {Cano}}, \bibinfo {author}
  {\bibfnamefont {K.}~\bibnamefont {Delaney}}, \bibinfo {author} {\bibfnamefont
  {Y.}~\bibnamefont {Kumagai}}, \bibinfo {author} {\bibfnamefont
  {M.}~\bibnamefont {Mostovoy}}, \bibinfo {author} {\bibfnamefont {N.~A.}\
  \bibnamefont {Spaldin}}, \bibinfo {author} {\bibfnamefont {R.}~\bibnamefont
  {Ramesh}},\ and\ \bibinfo {author} {\bibfnamefont {M.}~\bibnamefont
  {Fiebig}},\ }\bibfield  {title} {\bibinfo {title} {Anisotropic conductance at
  improper ferroelectric domain walls},\ }\href
  {https://doi.org/10.1038/nmat3249} {\bibfield  {journal} {\bibinfo  {journal}
  {Nature Materials}\ }\textbf {\bibinfo {volume} {11}},\ \bibinfo {pages}
  {284} (\bibinfo {year} {2012})}\BibitemShut {NoStop}%
\bibitem [{\citenamefont {Catalan}\ \emph {et~al.}(2012)\citenamefont
  {Catalan}, \citenamefont {Seidel}, \citenamefont {Ramesh},\ and\
  \citenamefont {Scott}}]{Catalan2012}%
  \BibitemOpen
  \bibfield  {author} {\bibinfo {author} {\bibfnamefont {G.}~\bibnamefont
  {Catalan}}, \bibinfo {author} {\bibfnamefont {J.}~\bibnamefont {Seidel}},
  \bibinfo {author} {\bibfnamefont {R.}~\bibnamefont {Ramesh}},\ and\ \bibinfo
  {author} {\bibfnamefont {J.~F.}\ \bibnamefont {Scott}},\ }\bibfield  {title}
  {\bibinfo {title} {Domain wall nanoelectronics},\ }\href
  {https://doi.org/10.1103/revmodphys.84.119} {\bibfield  {journal} {\bibinfo
  {journal} {Reviews of Modern Physics}\ }\textbf {\bibinfo {volume} {84}},\
  \bibinfo {pages} {119} (\bibinfo {year} {2012})}\BibitemShut {NoStop}%
\bibitem [{\citenamefont {Meier}\ and\ \citenamefont
  {Selbach}(2021)}]{Meier2021}%
  \BibitemOpen
  \bibfield  {author} {\bibinfo {author} {\bibfnamefont {D.}~\bibnamefont
  {Meier}}\ and\ \bibinfo {author} {\bibfnamefont {S.~M.}\ \bibnamefont
  {Selbach}},\ }\bibfield  {title} {\bibinfo {title} {Ferroelectric domain
  walls for nanotechnology},\ }\bibfield  {journal} {\bibinfo  {journal}
  {Nature Reviews Materials}\ }\href
  {https://doi.org/10.1038/s41578-021-00375-z} {10.1038/s41578-021-00375-z}
  (\bibinfo {year} {2021})\BibitemShut {NoStop}%
\bibitem [{\citenamefont {Meier}\ \emph {et~al.}(2020)\citenamefont {Meier},
  \citenamefont {Seidel}, \citenamefont {Gregg},\ and\ \citenamefont
  {Ramesh}}]{Meier2020}%
  \BibitemOpen
  \bibfield  {author} {\bibinfo {author} {\bibfnamefont {D.}~\bibnamefont
  {Meier}}, \bibinfo {author} {\bibfnamefont {J.}~\bibnamefont {Seidel}},
  \bibinfo {author} {\bibfnamefont {M.}~\bibnamefont {Gregg}},\ and\ \bibinfo
  {author} {\bibfnamefont {R.}~\bibnamefont {Ramesh}},\ }\href
  {https://doi.org/10.1093/oso/9780198862499.001.0001} {\emph {\bibinfo {title}
  {Domain Walls}}}\ (\bibinfo  {publisher} {Oxford University Press},\ \bibinfo
  {year} {2020})\BibitemShut {NoStop}%
\bibitem [{\citenamefont {Seidel}\ \emph {et~al.}(2009)\citenamefont {Seidel},
  \citenamefont {Martin}, \citenamefont {He}, \citenamefont {Zhan},
  \citenamefont {Chu}, \citenamefont {Rother}, \citenamefont {Hawkridge},
  \citenamefont {Maksymovych}, \citenamefont {Yu}, \citenamefont {Gajek},
  \citenamefont {Balke}, \citenamefont {Kalinin}, \citenamefont {Gemming},
  \citenamefont {Wang}, \citenamefont {Catalan}, \citenamefont {Scott},
  \citenamefont {Spaldin}, \citenamefont {Orenstein},\ and\ \citenamefont
  {Ramesh}}]{Seidel2009}%
  \BibitemOpen
  \bibfield  {author} {\bibinfo {author} {\bibfnamefont {J.}~\bibnamefont
  {Seidel}}, \bibinfo {author} {\bibfnamefont {L.~W.}\ \bibnamefont {Martin}},
  \bibinfo {author} {\bibfnamefont {Q.}~\bibnamefont {He}}, \bibinfo {author}
  {\bibfnamefont {Q.}~\bibnamefont {Zhan}}, \bibinfo {author} {\bibfnamefont
  {Y.-H.}\ \bibnamefont {Chu}}, \bibinfo {author} {\bibfnamefont
  {A.}~\bibnamefont {Rother}}, \bibinfo {author} {\bibfnamefont {M.~E.}\
  \bibnamefont {Hawkridge}}, \bibinfo {author} {\bibfnamefont {P.}~\bibnamefont
  {Maksymovych}}, \bibinfo {author} {\bibfnamefont {P.}~\bibnamefont {Yu}},
  \bibinfo {author} {\bibfnamefont {M.}~\bibnamefont {Gajek}}, \bibinfo
  {author} {\bibfnamefont {N.}~\bibnamefont {Balke}}, \bibinfo {author}
  {\bibfnamefont {S.~V.}\ \bibnamefont {Kalinin}}, \bibinfo {author}
  {\bibfnamefont {S.}~\bibnamefont {Gemming}}, \bibinfo {author} {\bibfnamefont
  {F.}~\bibnamefont {Wang}}, \bibinfo {author} {\bibfnamefont {G.}~\bibnamefont
  {Catalan}}, \bibinfo {author} {\bibfnamefont {J.~F.}\ \bibnamefont {Scott}},
  \bibinfo {author} {\bibfnamefont {N.~A.}\ \bibnamefont {Spaldin}}, \bibinfo
  {author} {\bibfnamefont {J.}~\bibnamefont {Orenstein}},\ and\ \bibinfo
  {author} {\bibfnamefont {R.}~\bibnamefont {Ramesh}},\ }\bibfield  {title}
  {\bibinfo {title} {Conduction at domain walls in oxide multiferroics},\
  }\href {https://doi.org/10.1038/nmat2373} {\bibfield  {journal} {\bibinfo
  {journal} {Nature Materials}\ }\textbf {\bibinfo {volume} {8}},\ \bibinfo
  {pages} {229} (\bibinfo {year} {2009})}\BibitemShut {NoStop}%
\bibitem [{\citenamefont {Soergel}(2011)}]{Soergel2011}%
  \BibitemOpen
  \bibfield  {author} {\bibinfo {author} {\bibfnamefont {E.}~\bibnamefont
  {Soergel}},\ }\bibfield  {title} {\bibinfo {title} {Piezoresponse force
  microscopy ({PFM})},\ }\href {https://doi.org/10.1088/0022-3727/44/46/464003}
  {\bibfield  {journal} {\bibinfo  {journal} {Journal of Physics D: Applied
  Physics}\ }\textbf {\bibinfo {volume} {44}},\ \bibinfo {pages} {464003}
  (\bibinfo {year} {2011})}\BibitemShut {NoStop}%
\bibitem [{\citenamefont {Guy}\ \emph {et~al.}(2021)\citenamefont {Guy},
  \citenamefont {Cochard}, \citenamefont {Aguado-Puente}, \citenamefont
  {Soergel}, \citenamefont {Whatmore}, \citenamefont {Conroy}, \citenamefont
  {Moore}, \citenamefont {Courtney}, \citenamefont {Harvey}, \citenamefont
  {Bangert}, \citenamefont {Kumar}, \citenamefont {McQuaid},\ and\
  \citenamefont {Gregg}}]{Guy2021}%
  \BibitemOpen
  \bibfield  {author} {\bibinfo {author} {\bibfnamefont {J.~G.~M.}\
  \bibnamefont {Guy}}, \bibinfo {author} {\bibfnamefont {C.}~\bibnamefont
  {Cochard}}, \bibinfo {author} {\bibfnamefont {P.}~\bibnamefont
  {Aguado-Puente}}, \bibinfo {author} {\bibfnamefont {E.}~\bibnamefont
  {Soergel}}, \bibinfo {author} {\bibfnamefont {R.~W.}\ \bibnamefont
  {Whatmore}}, \bibinfo {author} {\bibfnamefont {M.}~\bibnamefont {Conroy}},
  \bibinfo {author} {\bibfnamefont {K.}~\bibnamefont {Moore}}, \bibinfo
  {author} {\bibfnamefont {E.}~\bibnamefont {Courtney}}, \bibinfo {author}
  {\bibfnamefont {A.}~\bibnamefont {Harvey}}, \bibinfo {author} {\bibfnamefont
  {U.}~\bibnamefont {Bangert}}, \bibinfo {author} {\bibfnamefont
  {A.}~\bibnamefont {Kumar}}, \bibinfo {author} {\bibfnamefont {R.~G.~P.}\
  \bibnamefont {McQuaid}},\ and\ \bibinfo {author} {\bibfnamefont {J.~M.}\
  \bibnamefont {Gregg}},\ }\bibfield  {title} {\bibinfo {title} {Anomalous
  motion of charged domain walls and associated negative capacitance in
  copper{\textendash}chlorine boracite},\ }\href
  {https://doi.org/10.1002/adma.202008068} {\bibfield  {journal} {\bibinfo
  {journal} {Advanced Materials}\ }\textbf {\bibinfo {volume} {33}},\ \bibinfo
  {pages} {2008068} (\bibinfo {year} {2021})}\BibitemShut {NoStop}%
\bibitem [{\citenamefont {Schulthei{\ss}}\ \emph {et~al.}(2020)\citenamefont
  {Schulthei{\ss}}, \citenamefont {Schaab}, \citenamefont {Sm{\aa}br{\aa}ten},
  \citenamefont {Skj{\ae}rv{\o}}, \citenamefont {Bourret}, \citenamefont {Yan},
  \citenamefont {Selbach},\ and\ \citenamefont {Meier}}]{Schultheis2020}%
  \BibitemOpen
  \bibfield  {author} {\bibinfo {author} {\bibfnamefont {J.}~\bibnamefont
  {Schulthei{\ss}}}, \bibinfo {author} {\bibfnamefont {J.}~\bibnamefont
  {Schaab}}, \bibinfo {author} {\bibfnamefont {D.~R.}\ \bibnamefont
  {Sm{\aa}br{\aa}ten}}, \bibinfo {author} {\bibfnamefont {S.~H.}\ \bibnamefont
  {Skj{\ae}rv{\o}}}, \bibinfo {author} {\bibfnamefont {E.}~\bibnamefont
  {Bourret}}, \bibinfo {author} {\bibfnamefont {Z.}~\bibnamefont {Yan}},
  \bibinfo {author} {\bibfnamefont {S.~M.}\ \bibnamefont {Selbach}},\ and\
  \bibinfo {author} {\bibfnamefont {D.}~\bibnamefont {Meier}},\ }\bibfield
  {title} {\bibinfo {title} {Intrinsic and extrinsic conduction contributions
  at nominally neutral domain walls in hexagonal manganites},\ }\href
  {https://doi.org/10.1063/5.0009185} {\bibfield  {journal} {\bibinfo
  {journal} {Applied Physics Letters}\ }\textbf {\bibinfo {volume} {116}},\
  \bibinfo {pages} {262903} (\bibinfo {year} {2020})}\BibitemShut {NoStop}%
\bibitem [{\citenamefont {Jia}\ \emph {et~al.}(2007)\citenamefont {Jia},
  \citenamefont {Mi}, \citenamefont {Urban}, \citenamefont {Vrejoiu},
  \citenamefont {Alexe},\ and\ \citenamefont {Hesse}}]{Jia2007}%
  \BibitemOpen
  \bibfield  {author} {\bibinfo {author} {\bibfnamefont {C.-L.}\ \bibnamefont
  {Jia}}, \bibinfo {author} {\bibfnamefont {S.-B.}\ \bibnamefont {Mi}},
  \bibinfo {author} {\bibfnamefont {K.}~\bibnamefont {Urban}}, \bibinfo
  {author} {\bibfnamefont {I.}~\bibnamefont {Vrejoiu}}, \bibinfo {author}
  {\bibfnamefont {M.}~\bibnamefont {Alexe}},\ and\ \bibinfo {author}
  {\bibfnamefont {D.}~\bibnamefont {Hesse}},\ }\bibfield  {title} {\bibinfo
  {title} {Atomic-scale study of electric dipoles near charged and uncharged
  domain walls in ferroelectric films},\ }\href
  {https://doi.org/10.1038/nmat2080} {\bibfield  {journal} {\bibinfo  {journal}
  {Nature Materials}\ }\textbf {\bibinfo {volume} {7}},\ \bibinfo {pages} {57}
  (\bibinfo {year} {2007})}\BibitemShut {NoStop}%
\bibitem [{\citenamefont {Pawlik}\ \emph {et~al.}(2017)\citenamefont {Pawlik},
  \citenamefont {Kämpfe}, \citenamefont {Hau{\ss}mann}, \citenamefont {Woike},
  \citenamefont {Treske}, \citenamefont {Knupfer}, \citenamefont {Büchner},
  \citenamefont {Soergel}, \citenamefont {Streubel}, \citenamefont {Koitzsch},\
  and\ \citenamefont {Eng}}]{Pawlik2017}%
  \BibitemOpen
  \bibfield  {author} {\bibinfo {author} {\bibfnamefont {A.-S.}\ \bibnamefont
  {Pawlik}}, \bibinfo {author} {\bibfnamefont {T.}~\bibnamefont {Kämpfe}},
  \bibinfo {author} {\bibfnamefont {A.}~\bibnamefont {Hau{\ss}mann}}, \bibinfo
  {author} {\bibfnamefont {T.}~\bibnamefont {Woike}}, \bibinfo {author}
  {\bibfnamefont {U.}~\bibnamefont {Treske}}, \bibinfo {author} {\bibfnamefont
  {M.}~\bibnamefont {Knupfer}}, \bibinfo {author} {\bibfnamefont
  {B.}~\bibnamefont {Büchner}}, \bibinfo {author} {\bibfnamefont
  {E.}~\bibnamefont {Soergel}}, \bibinfo {author} {\bibfnamefont
  {R.}~\bibnamefont {Streubel}}, \bibinfo {author} {\bibfnamefont
  {A.}~\bibnamefont {Koitzsch}},\ and\ \bibinfo {author} {\bibfnamefont
  {L.~M.}\ \bibnamefont {Eng}},\ }\bibfield  {title} {\bibinfo {title}
  {Polarization driven conductance variations at charged ferroelectric domain
  walls},\ }\href {https://doi.org/10.1039/c7nr00217c} {\bibfield  {journal}
  {\bibinfo  {journal} {Nanoscale}\ }\textbf {\bibinfo {volume} {9}},\ \bibinfo
  {pages} {10933} (\bibinfo {year} {2017})}\BibitemShut {NoStop}%
\bibitem [{\citenamefont {Schaab}\ \emph {et~al.}(2014)\citenamefont {Schaab},
  \citenamefont {Krug}, \citenamefont {Nickel}, \citenamefont {Gottlob},
  \citenamefont {Do{\u{g}}anay}, \citenamefont {Cano}, \citenamefont
  {Hentschel}, \citenamefont {Yan}, \citenamefont {Bourret}, \citenamefont
  {Schneider}, \citenamefont {Ramesh},\ and\ \citenamefont
  {Meier}}]{Schaab2014}%
  \BibitemOpen
  \bibfield  {author} {\bibinfo {author} {\bibfnamefont {J.}~\bibnamefont
  {Schaab}}, \bibinfo {author} {\bibfnamefont {I.~P.}\ \bibnamefont {Krug}},
  \bibinfo {author} {\bibfnamefont {F.}~\bibnamefont {Nickel}}, \bibinfo
  {author} {\bibfnamefont {D.~M.}\ \bibnamefont {Gottlob}}, \bibinfo {author}
  {\bibfnamefont {H.}~\bibnamefont {Do{\u{g}}anay}}, \bibinfo {author}
  {\bibfnamefont {A.}~\bibnamefont {Cano}}, \bibinfo {author} {\bibfnamefont
  {M.}~\bibnamefont {Hentschel}}, \bibinfo {author} {\bibfnamefont
  {Z.}~\bibnamefont {Yan}}, \bibinfo {author} {\bibfnamefont {E.}~\bibnamefont
  {Bourret}}, \bibinfo {author} {\bibfnamefont {C.~M.}\ \bibnamefont
  {Schneider}}, \bibinfo {author} {\bibfnamefont {R.}~\bibnamefont {Ramesh}},\
  and\ \bibinfo {author} {\bibfnamefont {D.}~\bibnamefont {Meier}},\ }\bibfield
   {title} {\bibinfo {title} {Imaging and characterization of conducting
  ferroelectric domain walls by photoemission electron microscopy},\ }\href
  {https://doi.org/10.1063/1.4879260} {\bibfield  {journal} {\bibinfo
  {journal} {Applied Physics Letters}\ }\textbf {\bibinfo {volume} {104}},\
  \bibinfo {pages} {232904} (\bibinfo {year} {2014})}\BibitemShut {NoStop}%
\bibitem [{\citenamefont {Becher}\ \emph {et~al.}(2015)\citenamefont {Becher},
  \citenamefont {Maurel}, \citenamefont {Aschauer}, \citenamefont {Lilienblum},
  \citenamefont {Mag{\'{e}}n}, \citenamefont {Meier}, \citenamefont
  {Langenberg}, \citenamefont {Trassin}, \citenamefont {Blasco}, \citenamefont
  {Krug}, \citenamefont {Algarabel}, \citenamefont {Spaldin}, \citenamefont
  {Pardo},\ and\ \citenamefont {Fiebig}}]{Becher2015}%
  \BibitemOpen
  \bibfield  {author} {\bibinfo {author} {\bibfnamefont {C.}~\bibnamefont
  {Becher}}, \bibinfo {author} {\bibfnamefont {L.}~\bibnamefont {Maurel}},
  \bibinfo {author} {\bibfnamefont {U.}~\bibnamefont {Aschauer}}, \bibinfo
  {author} {\bibfnamefont {M.}~\bibnamefont {Lilienblum}}, \bibinfo {author}
  {\bibfnamefont {C.}~\bibnamefont {Mag{\'{e}}n}}, \bibinfo {author}
  {\bibfnamefont {D.}~\bibnamefont {Meier}}, \bibinfo {author} {\bibfnamefont
  {E.}~\bibnamefont {Langenberg}}, \bibinfo {author} {\bibfnamefont
  {M.}~\bibnamefont {Trassin}}, \bibinfo {author} {\bibfnamefont
  {J.}~\bibnamefont {Blasco}}, \bibinfo {author} {\bibfnamefont {I.~P.}\
  \bibnamefont {Krug}}, \bibinfo {author} {\bibfnamefont {P.~A.}\ \bibnamefont
  {Algarabel}}, \bibinfo {author} {\bibfnamefont {N.~A.}\ \bibnamefont
  {Spaldin}}, \bibinfo {author} {\bibfnamefont {J.~A.}\ \bibnamefont {Pardo}},\
  and\ \bibinfo {author} {\bibfnamefont {M.}~\bibnamefont {Fiebig}},\
  }\bibfield  {title} {\bibinfo {title} {Strain-induced coupling of electrical
  polarization and structural defects in {SrMnO}3 films},\ }\href
  {https://doi.org/10.1038/nnano.2015.108} {\bibfield  {journal} {\bibinfo
  {journal} {Nature Nanotechnology}\ }\textbf {\bibinfo {volume} {10}},\
  \bibinfo {pages} {661} (\bibinfo {year} {2015})}\BibitemShut {NoStop}%
\bibitem [{\citenamefont {Nataf}\ \emph {et~al.}(2016)\citenamefont {Nataf},
  \citenamefont {Grysan}, \citenamefont {Guennou}, \citenamefont {Kreisel},
  \citenamefont {Martinotti}, \citenamefont {Rountree}, \citenamefont
  {Mathieu},\ and\ \citenamefont {Barrett}}]{Nataf2016}%
  \BibitemOpen
  \bibfield  {author} {\bibinfo {author} {\bibfnamefont {G.~F.}\ \bibnamefont
  {Nataf}}, \bibinfo {author} {\bibfnamefont {P.}~\bibnamefont {Grysan}},
  \bibinfo {author} {\bibfnamefont {M.}~\bibnamefont {Guennou}}, \bibinfo
  {author} {\bibfnamefont {J.}~\bibnamefont {Kreisel}}, \bibinfo {author}
  {\bibfnamefont {D.}~\bibnamefont {Martinotti}}, \bibinfo {author}
  {\bibfnamefont {C.~L.}\ \bibnamefont {Rountree}}, \bibinfo {author}
  {\bibfnamefont {C.}~\bibnamefont {Mathieu}},\ and\ \bibinfo {author}
  {\bibfnamefont {N.}~\bibnamefont {Barrett}},\ }\bibfield  {title} {\bibinfo
  {title} {Low energy electron imaging of domains and domain walls in
  magnesium-doped lithium niobate},\ }\bibfield  {journal} {\bibinfo  {journal}
  {Scientific Reports}\ }\textbf {\bibinfo {volume} {6}},\ \href
  {https://doi.org/10.1038/srep33098} {10.1038/srep33098} (\bibinfo {year}
  {2016})\BibitemShut {NoStop}%
\bibitem [{\citenamefont {Hunnestad}\ \emph {et~al.}(2020)\citenamefont
  {Hunnestad}, \citenamefont {Roede}, \citenamefont {van Helvoort},\ and\
  \citenamefont {Meier}}]{Hunnestad2020}%
  \BibitemOpen
  \bibfield  {author} {\bibinfo {author} {\bibfnamefont {K.~A.}\ \bibnamefont
  {Hunnestad}}, \bibinfo {author} {\bibfnamefont {E.~D.}\ \bibnamefont
  {Roede}}, \bibinfo {author} {\bibfnamefont {A.~T.~J.}\ \bibnamefont {van
  Helvoort}},\ and\ \bibinfo {author} {\bibfnamefont {D.}~\bibnamefont
  {Meier}},\ }\bibfield  {title} {\bibinfo {title} {Characterization of
  ferroelectric domain walls by scanning electron microscopy},\ }\href
  {https://doi.org/10.1063/5.0029284} {\bibfield  {journal} {\bibinfo
  {journal} {Journal of Applied Physics}\ }\textbf {\bibinfo {volume} {128}},\
  \bibinfo {pages} {191102} (\bibinfo {year} {2020})}\BibitemShut {NoStop}%
\bibitem [{\citenamefont {Farokhipoor}\ and\ \citenamefont
  {Noheda}(2011)}]{Farokhipoor2011}%
  \BibitemOpen
  \bibfield  {author} {\bibinfo {author} {\bibfnamefont {S.}~\bibnamefont
  {Farokhipoor}}\ and\ \bibinfo {author} {\bibfnamefont {B.}~\bibnamefont
  {Noheda}},\ }\bibfield  {title} {\bibinfo {title} {Conduction through
  71{\textdegree} domain walls in \ce{BiFeO3} thin films},\ }\bibfield
  {journal} {\bibinfo  {journal} {Physical Review Letters}\ }\textbf {\bibinfo
  {volume} {107}},\ \href {https://doi.org/10.1103/physrevlett.107.127601}
  {10.1103/physrevlett.107.127601} (\bibinfo {year} {2011})\BibitemShut
  {NoStop}%
\bibitem [{\citenamefont {Guyonnet}\ \emph {et~al.}(2011)\citenamefont
  {Guyonnet}, \citenamefont {Gaponenko}, \citenamefont {Gariglio},\ and\
  \citenamefont {Paruch}}]{Guyonnet2011}%
  \BibitemOpen
  \bibfield  {author} {\bibinfo {author} {\bibfnamefont {J.}~\bibnamefont
  {Guyonnet}}, \bibinfo {author} {\bibfnamefont {I.}~\bibnamefont {Gaponenko}},
  \bibinfo {author} {\bibfnamefont {S.}~\bibnamefont {Gariglio}},\ and\
  \bibinfo {author} {\bibfnamefont {P.}~\bibnamefont {Paruch}},\ }\bibfield
  {title} {\bibinfo {title} {Conduction at domain walls in insulating
  \ce{Pb(Zr_{0.2}Ti_{0.8})O3} thin films},\ }\href
  {https://doi.org/10.1002/adma.201102254} {\bibfield  {journal} {\bibinfo
  {journal} {Advanced Materials}\ }\textbf {\bibinfo {volume} {23}},\ \bibinfo
  {pages} {5377} (\bibinfo {year} {2011})}\BibitemShut {NoStop}%
\bibitem [{\citenamefont {Sluka}\ \emph {et~al.}(2013)\citenamefont {Sluka},
  \citenamefont {Tagantsev}, \citenamefont {Bednyakov},\ and\ \citenamefont
  {Setter}}]{Sluka2013}%
  \BibitemOpen
  \bibfield  {author} {\bibinfo {author} {\bibfnamefont {T.}~\bibnamefont
  {Sluka}}, \bibinfo {author} {\bibfnamefont {A.~K.}\ \bibnamefont
  {Tagantsev}}, \bibinfo {author} {\bibfnamefont {P.}~\bibnamefont
  {Bednyakov}},\ and\ \bibinfo {author} {\bibfnamefont {N.}~\bibnamefont
  {Setter}},\ }\bibfield  {title} {\bibinfo {title} {Free-electron gas at
  charged domain walls in insulating \ce{BaTiO3}},\ }\bibfield  {journal}
  {\bibinfo  {journal} {Nature Communications}\ }\textbf {\bibinfo {volume}
  {4}},\ \href {https://doi.org/10.1038/ncomms2839} {10.1038/ncomms2839}
  (\bibinfo {year} {2013})\BibitemShut {NoStop}%
\bibitem [{\citenamefont {Schröder}\ \emph {et~al.}(2012)\citenamefont
  {Schröder}, \citenamefont {Hau{\ss}mann}, \citenamefont {Thiessen},
  \citenamefont {Soergel}, \citenamefont {Woike},\ and\ \citenamefont
  {Eng}}]{Schroeder2012}%
  \BibitemOpen
  \bibfield  {author} {\bibinfo {author} {\bibfnamefont {M.}~\bibnamefont
  {Schröder}}, \bibinfo {author} {\bibfnamefont {A.}~\bibnamefont
  {Hau{\ss}mann}}, \bibinfo {author} {\bibfnamefont {A.}~\bibnamefont
  {Thiessen}}, \bibinfo {author} {\bibfnamefont {E.}~\bibnamefont {Soergel}},
  \bibinfo {author} {\bibfnamefont {T.}~\bibnamefont {Woike}},\ and\ \bibinfo
  {author} {\bibfnamefont {L.~M.}\ \bibnamefont {Eng}},\ }\bibfield  {title}
  {\bibinfo {title} {Conducting domain walls in lithium niobate single
  crystals},\ }\href {https://doi.org/10.1002/adfm.201201174} {\bibfield
  {journal} {\bibinfo  {journal} {Advanced Functional Materials}\ }\textbf
  {\bibinfo {volume} {22}},\ \bibinfo {pages} {3936} (\bibinfo {year}
  {2012})}\BibitemShut {NoStop}%
\bibitem [{\citenamefont {Mosberg}\ \emph {et~al.}(2019)\citenamefont
  {Mosberg}, \citenamefont {Roede}, \citenamefont {Evans}, \citenamefont
  {Holstad}, \citenamefont {Bourret}, \citenamefont {Yan}, \citenamefont {van
  Helvoort},\ and\ \citenamefont {Meier}}]{Mosberg2019}%
  \BibitemOpen
  \bibfield  {author} {\bibinfo {author} {\bibfnamefont {A.~B.}\ \bibnamefont
  {Mosberg}}, \bibinfo {author} {\bibfnamefont {E.~D.}\ \bibnamefont {Roede}},
  \bibinfo {author} {\bibfnamefont {D.~M.}\ \bibnamefont {Evans}}, \bibinfo
  {author} {\bibfnamefont {T.~S.}\ \bibnamefont {Holstad}}, \bibinfo {author}
  {\bibfnamefont {E.}~\bibnamefont {Bourret}}, \bibinfo {author} {\bibfnamefont
  {Z.}~\bibnamefont {Yan}}, \bibinfo {author} {\bibfnamefont {A.~T.~J.}\
  \bibnamefont {van Helvoort}},\ and\ \bibinfo {author} {\bibfnamefont
  {D.}~\bibnamefont {Meier}},\ }\bibfield  {title} {\bibinfo {title} {{FIB}
  lift-out of conducting ferroelectric domain walls in hexagonal manganites},\
  }\href {https://doi.org/10.1063/1.5115465} {\bibfield  {journal} {\bibinfo
  {journal} {Applied Physics Letters}\ }\textbf {\bibinfo {volume} {115}},\
  \bibinfo {pages} {122901} (\bibinfo {year} {2019})}\BibitemShut {NoStop}%
\bibitem [{\citenamefont {Tselev}\ \emph {et~al.}(2016)\citenamefont {Tselev},
  \citenamefont {Yu}, \citenamefont {Cao}, \citenamefont {Dedon}, \citenamefont
  {Martin}, \citenamefont {Kalinin},\ and\ \citenamefont
  {Maksymovych}}]{Tselev2016}%
  \BibitemOpen
  \bibfield  {author} {\bibinfo {author} {\bibfnamefont {A.}~\bibnamefont
  {Tselev}}, \bibinfo {author} {\bibfnamefont {P.}~\bibnamefont {Yu}}, \bibinfo
  {author} {\bibfnamefont {Y.}~\bibnamefont {Cao}}, \bibinfo {author}
  {\bibfnamefont {L.~R.}\ \bibnamefont {Dedon}}, \bibinfo {author}
  {\bibfnamefont {L.~W.}\ \bibnamefont {Martin}}, \bibinfo {author}
  {\bibfnamefont {S.~V.}\ \bibnamefont {Kalinin}},\ and\ \bibinfo {author}
  {\bibfnamefont {P.}~\bibnamefont {Maksymovych}},\ }\bibfield  {title}
  {\bibinfo {title} {Microwave a.c. conductivity of domain walls in
  ferroelectric thin films},\ }\bibfield  {journal} {\bibinfo  {journal}
  {Nature Communications}\ }\textbf {\bibinfo {volume} {7}},\ \href
  {https://doi.org/10.1038/ncomms11630} {10.1038/ncomms11630} (\bibinfo {year}
  {2016})\BibitemShut {NoStop}%
\bibitem [{\citenamefont {Maksymovych}\ \emph {et~al.}(2011)\citenamefont
  {Maksymovych}, \citenamefont {Morozovska}, \citenamefont {Yu}, \citenamefont
  {Eliseev}, \citenamefont {Chu}, \citenamefont {Ramesh}, \citenamefont
  {Baddorf},\ and\ \citenamefont {Kalinin}}]{Maksymovych2011}%
  \BibitemOpen
  \bibfield  {author} {\bibinfo {author} {\bibfnamefont {P.}~\bibnamefont
  {Maksymovych}}, \bibinfo {author} {\bibfnamefont {A.~N.}\ \bibnamefont
  {Morozovska}}, \bibinfo {author} {\bibfnamefont {P.}~\bibnamefont {Yu}},
  \bibinfo {author} {\bibfnamefont {E.~A.}\ \bibnamefont {Eliseev}}, \bibinfo
  {author} {\bibfnamefont {Y.-H.}\ \bibnamefont {Chu}}, \bibinfo {author}
  {\bibfnamefont {R.}~\bibnamefont {Ramesh}}, \bibinfo {author} {\bibfnamefont
  {A.~P.}\ \bibnamefont {Baddorf}},\ and\ \bibinfo {author} {\bibfnamefont
  {S.~V.}\ \bibnamefont {Kalinin}},\ }\bibfield  {title} {\bibinfo {title}
  {Tunable metallic conductance in ferroelectric nanodomains},\ }\href
  {https://doi.org/10.1021/nl203349b} {\bibfield  {journal} {\bibinfo
  {journal} {Nano Letters}\ }\textbf {\bibinfo {volume} {12}},\ \bibinfo
  {pages} {209} (\bibinfo {year} {2011})}\BibitemShut {NoStop}%
\bibitem [{\citenamefont {Gureev}\ \emph {et~al.}(2011)\citenamefont {Gureev},
  \citenamefont {Tagantsev},\ and\ \citenamefont {Setter}}]{Gureev2011}%
  \BibitemOpen
  \bibfield  {author} {\bibinfo {author} {\bibfnamefont {M.~Y.}\ \bibnamefont
  {Gureev}}, \bibinfo {author} {\bibfnamefont {A.~K.}\ \bibnamefont
  {Tagantsev}},\ and\ \bibinfo {author} {\bibfnamefont {N.}~\bibnamefont
  {Setter}},\ }\bibfield  {title} {\bibinfo {title} {Head-to-head and
  tail-to-tail 180{\textdegree} domain walls in an isolated ferroelectric},\
  }\bibfield  {journal} {\bibinfo  {journal} {Physical Review B}\ }\textbf
  {\bibinfo {volume} {83}},\ \href {https://doi.org/10.1103/physrevb.83.184104}
  {10.1103/physrevb.83.184104} (\bibinfo {year} {2011})\BibitemShut {NoStop}%
\bibitem [{\citenamefont {Gonnissen}\ \emph {et~al.}(2016)\citenamefont
  {Gonnissen}, \citenamefont {Batuk}, \citenamefont {Nataf}, \citenamefont
  {Jones}, \citenamefont {Abakumov}, \citenamefont {Aert}, \citenamefont
  {Schryvers},\ and\ \citenamefont {Salje}}]{Gonnissen2016}%
  \BibitemOpen
  \bibfield  {author} {\bibinfo {author} {\bibfnamefont {J.}~\bibnamefont
  {Gonnissen}}, \bibinfo {author} {\bibfnamefont {D.}~\bibnamefont {Batuk}},
  \bibinfo {author} {\bibfnamefont {G.~F.}\ \bibnamefont {Nataf}}, \bibinfo
  {author} {\bibfnamefont {L.}~\bibnamefont {Jones}}, \bibinfo {author}
  {\bibfnamefont {A.~M.}\ \bibnamefont {Abakumov}}, \bibinfo {author}
  {\bibfnamefont {S.~V.}\ \bibnamefont {Aert}}, \bibinfo {author}
  {\bibfnamefont {D.}~\bibnamefont {Schryvers}},\ and\ \bibinfo {author}
  {\bibfnamefont {E.~K.~H.}\ \bibnamefont {Salje}},\ }\bibfield  {title}
  {\bibinfo {title} {Direct observation of ferroelectric domain walls in
  \ce{LiNbO3}: Wall-meanders, kinks, and local electric charges},\ }\href
  {https://doi.org/10.1002/adfm.201603489} {\bibfield  {journal} {\bibinfo
  {journal} {Advanced Functional Materials}\ }\textbf {\bibinfo {volume}
  {26}},\ \bibinfo {pages} {7599} (\bibinfo {year} {2016})}\BibitemShut
  {NoStop}%
\bibitem [{\citenamefont {Zhang}\ \emph {et~al.}(2012)\citenamefont {Zhang},
  \citenamefont {Wang}, \citenamefont {Wei}, \citenamefont {Yu}, \citenamefont
  {Gu}, \citenamefont {Hirata}, \citenamefont {Chen}, \citenamefont {Jin},
  \citenamefont {Yao}, \citenamefont {Wang},\ and\ \citenamefont
  {Duan}}]{Zhang2012}%
  \BibitemOpen
  \bibfield  {author} {\bibinfo {author} {\bibfnamefont {Q.~H.}\ \bibnamefont
  {Zhang}}, \bibinfo {author} {\bibfnamefont {L.~J.}\ \bibnamefont {Wang}},
  \bibinfo {author} {\bibfnamefont {X.~K.}\ \bibnamefont {Wei}}, \bibinfo
  {author} {\bibfnamefont {R.~C.}\ \bibnamefont {Yu}}, \bibinfo {author}
  {\bibfnamefont {L.}~\bibnamefont {Gu}}, \bibinfo {author} {\bibfnamefont
  {A.}~\bibnamefont {Hirata}}, \bibinfo {author} {\bibfnamefont {M.~W.}\
  \bibnamefont {Chen}}, \bibinfo {author} {\bibfnamefont {C.~Q.}\ \bibnamefont
  {Jin}}, \bibinfo {author} {\bibfnamefont {Y.}~\bibnamefont {Yao}}, \bibinfo
  {author} {\bibfnamefont {Y.~G.}\ \bibnamefont {Wang}},\ and\ \bibinfo
  {author} {\bibfnamefont {X.~F.}\ \bibnamefont {Duan}},\ }\bibfield  {title}
  {\bibinfo {title} {Direct observation of interlocked domain walls in
  hexagonal \ce{RMnO3 (R=Tm, Lu)}},\ }\bibfield  {journal} {\bibinfo  {journal}
  {Physical Review B}\ }\textbf {\bibinfo {volume} {85}},\ \href
  {https://doi.org/10.1103/physrevb.85.020102} {10.1103/physrevb.85.020102}
  (\bibinfo {year} {2012})\BibitemShut {NoStop}%
\bibitem [{\citenamefont {Holtz}\ \emph {et~al.}(2017)\citenamefont {Holtz},
  \citenamefont {Shapovalov}, \citenamefont {Mundy}, \citenamefont {Chang},
  \citenamefont {Yan}, \citenamefont {Bourret}, \citenamefont {Muller},
  \citenamefont {Meier},\ and\ \citenamefont {Cano}}]{Holtz2017}%
  \BibitemOpen
  \bibfield  {author} {\bibinfo {author} {\bibfnamefont {M.~E.}\ \bibnamefont
  {Holtz}}, \bibinfo {author} {\bibfnamefont {K.}~\bibnamefont {Shapovalov}},
  \bibinfo {author} {\bibfnamefont {J.~A.}\ \bibnamefont {Mundy}}, \bibinfo
  {author} {\bibfnamefont {C.~S.}\ \bibnamefont {Chang}}, \bibinfo {author}
  {\bibfnamefont {Z.}~\bibnamefont {Yan}}, \bibinfo {author} {\bibfnamefont
  {E.}~\bibnamefont {Bourret}}, \bibinfo {author} {\bibfnamefont {D.~A.}\
  \bibnamefont {Muller}}, \bibinfo {author} {\bibfnamefont {D.}~\bibnamefont
  {Meier}},\ and\ \bibinfo {author} {\bibfnamefont {A.}~\bibnamefont {Cano}},\
  }\bibfield  {title} {\bibinfo {title} {Topological defects in hexagonal
  manganites: Inner structure and emergent electrostatics},\ }\href
  {https://doi.org/10.1021/acs.nanolett.7b01288} {\bibfield  {journal}
  {\bibinfo  {journal} {Nano Letters}\ }\textbf {\bibinfo {volume} {17}},\
  \bibinfo {pages} {5883} (\bibinfo {year} {2017})}\BibitemShut {NoStop}%
\bibitem [{\citenamefont {McConville}\ \emph {et~al.}(2020)\citenamefont
  {McConville}, \citenamefont {Lu}, \citenamefont {Wang}, \citenamefont {Tan},
  \citenamefont {Cochard}, \citenamefont {Conroy}, \citenamefont {Moore},
  \citenamefont {Harvey}, \citenamefont {Bangert}, \citenamefont {Chen},
  \citenamefont {Gruverman},\ and\ \citenamefont {Gregg}}]{McConville2020}%
  \BibitemOpen
  \bibfield  {author} {\bibinfo {author} {\bibfnamefont {J.~P.~V.}\
  \bibnamefont {McConville}}, \bibinfo {author} {\bibfnamefont
  {H.}~\bibnamefont {Lu}}, \bibinfo {author} {\bibfnamefont {B.}~\bibnamefont
  {Wang}}, \bibinfo {author} {\bibfnamefont {Y.}~\bibnamefont {Tan}}, \bibinfo
  {author} {\bibfnamefont {C.}~\bibnamefont {Cochard}}, \bibinfo {author}
  {\bibfnamefont {M.}~\bibnamefont {Conroy}}, \bibinfo {author} {\bibfnamefont
  {K.}~\bibnamefont {Moore}}, \bibinfo {author} {\bibfnamefont
  {A.}~\bibnamefont {Harvey}}, \bibinfo {author} {\bibfnamefont
  {U.}~\bibnamefont {Bangert}}, \bibinfo {author} {\bibfnamefont {L.-Q.}\
  \bibnamefont {Chen}}, \bibinfo {author} {\bibfnamefont {A.}~\bibnamefont
  {Gruverman}},\ and\ \bibinfo {author} {\bibfnamefont {J.~M.}\ \bibnamefont
  {Gregg}},\ }\bibfield  {title} {\bibinfo {title} {Ferroelectric domain wall
  memristor},\ }\href {https://doi.org/10.1002/adfm.202000109} {\bibfield
  {journal} {\bibinfo  {journal} {Advanced Functional Materials}\ }\textbf
  {\bibinfo {volume} {30}},\ \bibinfo {pages} {2000109} (\bibinfo {year}
  {2020})}\BibitemShut {NoStop}%
\bibitem [{\citenamefont {Zhang}\ \emph {et~al.}(2019)\citenamefont {Zhang},
  \citenamefont {Lu}, \citenamefont {Yan}, \citenamefont {Cheng}, \citenamefont
  {Xie}, \citenamefont {Aoki}, \citenamefont {Li}, \citenamefont {Heikes},
  \citenamefont {Lau}, \citenamefont {Schlom}, \citenamefont {Chen},
  \citenamefont {Gruverman},\ and\ \citenamefont {Pan}}]{Zhang2019}%
  \BibitemOpen
  \bibfield  {author} {\bibinfo {author} {\bibfnamefont {Y.}~\bibnamefont
  {Zhang}}, \bibinfo {author} {\bibfnamefont {H.}~\bibnamefont {Lu}}, \bibinfo
  {author} {\bibfnamefont {X.}~\bibnamefont {Yan}}, \bibinfo {author}
  {\bibfnamefont {X.}~\bibnamefont {Cheng}}, \bibinfo {author} {\bibfnamefont
  {L.}~\bibnamefont {Xie}}, \bibinfo {author} {\bibfnamefont {T.}~\bibnamefont
  {Aoki}}, \bibinfo {author} {\bibfnamefont {L.}~\bibnamefont {Li}}, \bibinfo
  {author} {\bibfnamefont {C.}~\bibnamefont {Heikes}}, \bibinfo {author}
  {\bibfnamefont {S.~P.}\ \bibnamefont {Lau}}, \bibinfo {author} {\bibfnamefont
  {D.~G.}\ \bibnamefont {Schlom}}, \bibinfo {author} {\bibfnamefont
  {L.}~\bibnamefont {Chen}}, \bibinfo {author} {\bibfnamefont {A.}~\bibnamefont
  {Gruverman}},\ and\ \bibinfo {author} {\bibfnamefont {X.}~\bibnamefont
  {Pan}},\ }\bibfield  {title} {\bibinfo {title} {Intrinsic conductance of
  domain walls in \ce{BiFeO3}},\ }\href
  {https://doi.org/10.1002/adma.201902099} {\bibfield  {journal} {\bibinfo
  {journal} {Advanced Materials}\ }\textbf {\bibinfo {volume} {31}},\ \bibinfo
  {pages} {1902099} (\bibinfo {year} {2019})}\BibitemShut {NoStop}%
\bibitem [{\citenamefont {Uesu}\ \emph {et~al.}(2004)\citenamefont {Uesu},
  \citenamefont {Shibata}, \citenamefont {Suzuki},\ and\ \citenamefont
  {Shimada}}]{Y.2004}%
  \BibitemOpen
  \bibfield  {author} {\bibinfo {author} {\bibfnamefont {Y.}~\bibnamefont
  {Uesu}}, \bibinfo {author} {\bibfnamefont {H.}~\bibnamefont {Shibata}},
  \bibinfo {author} {\bibfnamefont {S.}~\bibnamefont {Suzuki}},\ and\ \bibinfo
  {author} {\bibfnamefont {S.}~\bibnamefont {Shimada}},\ }\bibfield  {title}
  {\bibinfo {title} {3d images of inverted domain structure in \ce{LiNbO3}
  using shg interference microscope},\ }\href
  {https://doi.org/10.1080/00150190490457618} {\bibfield  {journal} {\bibinfo
  {journal} {Ferroelectrics}\ }\textbf {\bibinfo {volume} {304}},\ \bibinfo
  {pages} {99} (\bibinfo {year} {2004})}\BibitemShut {NoStop}%
\bibitem [{\citenamefont {Sheng}\ \emph {et~al.}(2010)\citenamefont {Sheng},
  \citenamefont {Best}, \citenamefont {Butt}, \citenamefont {Krolikowski},
  \citenamefont {Arie},\ and\ \citenamefont {Koynov}}]{Sheng2010}%
  \BibitemOpen
  \bibfield  {author} {\bibinfo {author} {\bibfnamefont {Y.}~\bibnamefont
  {Sheng}}, \bibinfo {author} {\bibfnamefont {A.}~\bibnamefont {Best}},
  \bibinfo {author} {\bibfnamefont {H.-J.}\ \bibnamefont {Butt}}, \bibinfo
  {author} {\bibfnamefont {W.}~\bibnamefont {Krolikowski}}, \bibinfo {author}
  {\bibfnamefont {A.}~\bibnamefont {Arie}},\ and\ \bibinfo {author}
  {\bibfnamefont {K.}~\bibnamefont {Koynov}},\ }\bibfield  {title} {\bibinfo
  {title} {Three-dimensional ferroelectric domain visualization by
  {\v{c}}erenkov-type second harmonic generation},\ }\href
  {https://doi.org/10.1364/oe.18.016539} {\bibfield  {journal} {\bibinfo
  {journal} {Optics Express}\ }\textbf {\bibinfo {volume} {18}},\ \bibinfo
  {pages} {16539} (\bibinfo {year} {2010})}\BibitemShut {NoStop}%
\bibitem [{\citenamefont {Kämpfe}\ \emph {et~al.}(2014)\citenamefont
  {Kämpfe}, \citenamefont {Reichenbach}, \citenamefont {Schröder},
  \citenamefont {Hau{\ss}mann}, \citenamefont {Eng}, \citenamefont {Woike},\
  and\ \citenamefont {Soergel}}]{Kaempfe2014}%
  \BibitemOpen
  \bibfield  {author} {\bibinfo {author} {\bibfnamefont {T.}~\bibnamefont
  {Kämpfe}}, \bibinfo {author} {\bibfnamefont {P.}~\bibnamefont
  {Reichenbach}}, \bibinfo {author} {\bibfnamefont {M.}~\bibnamefont
  {Schröder}}, \bibinfo {author} {\bibfnamefont {A.}~\bibnamefont
  {Hau{\ss}mann}}, \bibinfo {author} {\bibfnamefont {L.~M.}\ \bibnamefont
  {Eng}}, \bibinfo {author} {\bibfnamefont {T.}~\bibnamefont {Woike}},\ and\
  \bibinfo {author} {\bibfnamefont {E.}~\bibnamefont {Soergel}},\ }\bibfield
  {title} {\bibinfo {title} {Optical three-dimensional profiling of charged
  domain walls in ferroelectrics by cherenkov second-harmonic generation},\
  }\bibfield  {journal} {\bibinfo  {journal} {Physical Review B}\ }\textbf
  {\bibinfo {volume} {89}},\ \href {https://doi.org/10.1103/physrevb.89.035314}
  {10.1103/physrevb.89.035314} (\bibinfo {year} {2014})\BibitemShut {NoStop}%
\bibitem [{\citenamefont {Godau}\ \emph {et~al.}(2017)\citenamefont {Godau},
  \citenamefont {Kämpfe}, \citenamefont {Thiessen}, \citenamefont {Eng},\ and\
  \citenamefont {Hau{\ss}mann}}]{Godau2017}%
  \BibitemOpen
  \bibfield  {author} {\bibinfo {author} {\bibfnamefont {C.}~\bibnamefont
  {Godau}}, \bibinfo {author} {\bibfnamefont {T.}~\bibnamefont {Kämpfe}},
  \bibinfo {author} {\bibfnamefont {A.}~\bibnamefont {Thiessen}}, \bibinfo
  {author} {\bibfnamefont {L.~M.}\ \bibnamefont {Eng}},\ and\ \bibinfo {author}
  {\bibfnamefont {A.}~\bibnamefont {Hau{\ss}mann}},\ }\bibfield  {title}
  {\bibinfo {title} {Enhancing the domain wall conductivity in lithium niobate
  single crystals},\ }\href {https://doi.org/10.1021/acsnano.7b01199}
  {\bibfield  {journal} {\bibinfo  {journal} {{ACS} Nano}\ }\textbf {\bibinfo
  {volume} {11}},\ \bibinfo {pages} {4816} (\bibinfo {year}
  {2017})}\BibitemShut {NoStop}%
\bibitem [{\citenamefont {Cherifi-Hertel}\ \emph {et~al.}(2017)\citenamefont
  {Cherifi-Hertel}, \citenamefont {Bulou}, \citenamefont {Hertel},
  \citenamefont {Taupier}, \citenamefont {Dorkenoo}, \citenamefont {Andreas},
  \citenamefont {Guyonnet}, \citenamefont {Gaponenko}, \citenamefont {Gallo},\
  and\ \citenamefont {Paruch}}]{CherifiHertel2017}%
  \BibitemOpen
  \bibfield  {author} {\bibinfo {author} {\bibfnamefont {S.}~\bibnamefont
  {Cherifi-Hertel}}, \bibinfo {author} {\bibfnamefont {H.}~\bibnamefont
  {Bulou}}, \bibinfo {author} {\bibfnamefont {R.}~\bibnamefont {Hertel}},
  \bibinfo {author} {\bibfnamefont {G.}~\bibnamefont {Taupier}}, \bibinfo
  {author} {\bibfnamefont {K.~D.}\ \bibnamefont {Dorkenoo}}, \bibinfo {author}
  {\bibfnamefont {C.}~\bibnamefont {Andreas}}, \bibinfo {author} {\bibfnamefont
  {J.}~\bibnamefont {Guyonnet}}, \bibinfo {author} {\bibfnamefont
  {I.}~\bibnamefont {Gaponenko}}, \bibinfo {author} {\bibfnamefont
  {K.}~\bibnamefont {Gallo}},\ and\ \bibinfo {author} {\bibfnamefont
  {P.}~\bibnamefont {Paruch}},\ }\bibfield  {title} {\bibinfo {title}
  {Non-ising and chiral ferroelectric domain walls revealed by nonlinear
  optical microscopy},\ }\bibfield  {journal} {\bibinfo  {journal} {Nature
  Communications}\ }\textbf {\bibinfo {volume} {8}},\ \href
  {https://doi.org/10.1038/ncomms15768} {10.1038/ncomms15768} (\bibinfo {year}
  {2017})\BibitemShut {NoStop}%
\bibitem [{\citenamefont {Uesu}\ \emph {et~al.}(2007)\citenamefont {Uesu},
  \citenamefont {Yokota}, \citenamefont {Kawado}, \citenamefont {Kaneshiro},
  \citenamefont {Kurimura},\ and\ \citenamefont {Kato}}]{Uesu2007}%
  \BibitemOpen
  \bibfield  {author} {\bibinfo {author} {\bibfnamefont {Y.}~\bibnamefont
  {Uesu}}, \bibinfo {author} {\bibfnamefont {H.}~\bibnamefont {Yokota}},
  \bibinfo {author} {\bibfnamefont {S.}~\bibnamefont {Kawado}}, \bibinfo
  {author} {\bibfnamefont {J.}~\bibnamefont {Kaneshiro}}, \bibinfo {author}
  {\bibfnamefont {S.}~\bibnamefont {Kurimura}},\ and\ \bibinfo {author}
  {\bibfnamefont {N.}~\bibnamefont {Kato}},\ }\bibfield  {title} {\bibinfo
  {title} {Three-dimensional observations of periodically poled domains in a
  \ce{LiTaO3} quasiphase matching crystal by second harmonic generation
  tomography},\ }\href {https://doi.org/10.1063/1.2786589} {\bibfield
  {journal} {\bibinfo  {journal} {Applied Physics Letters}\ }\textbf {\bibinfo
  {volume} {91}},\ \bibinfo {pages} {182904} (\bibinfo {year}
  {2007})}\BibitemShut {NoStop}%
\bibitem [{\citenamefont {Yan}\ \emph {et~al.}(2015)\citenamefont {Yan},
  \citenamefont {Meier}, \citenamefont {Schaab}, \citenamefont {Ramesh},
  \citenamefont {Samulon},\ and\ \citenamefont {Bourret}}]{Yan2015}%
  \BibitemOpen
  \bibfield  {author} {\bibinfo {author} {\bibfnamefont {Z.}~\bibnamefont
  {Yan}}, \bibinfo {author} {\bibfnamefont {D.}~\bibnamefont {Meier}}, \bibinfo
  {author} {\bibfnamefont {J.}~\bibnamefont {Schaab}}, \bibinfo {author}
  {\bibfnamefont {R.}~\bibnamefont {Ramesh}}, \bibinfo {author} {\bibfnamefont
  {E.}~\bibnamefont {Samulon}},\ and\ \bibinfo {author} {\bibfnamefont
  {E.}~\bibnamefont {Bourret}},\ }\bibfield  {title} {\bibinfo {title} {Growth
  of high-quality hexagonal \ce{ErMnO3} single crystals by the pressurized
  floating-zone method},\ }\href
  {https://doi.org/10.1016/j.jcrysgro.2014.10.006} {\bibfield  {journal}
  {\bibinfo  {journal} {Journal of Crystal Growth}\ }\textbf {\bibinfo {volume}
  {409}},\ \bibinfo {pages} {75} (\bibinfo {year} {2015})}\BibitemShut
  {NoStop}%
\bibitem [{\citenamefont {Chae}\ \emph {et~al.}(2012)\citenamefont {Chae},
  \citenamefont {Lee}, \citenamefont {Horibe}, \citenamefont {Tanimura},
  \citenamefont {Mori}, \citenamefont {Gao}, \citenamefont {Carr},\ and\
  \citenamefont {Cheong}}]{Chae2012}%
  \BibitemOpen
  \bibfield  {author} {\bibinfo {author} {\bibfnamefont {S.~C.}\ \bibnamefont
  {Chae}}, \bibinfo {author} {\bibfnamefont {N.}~\bibnamefont {Lee}}, \bibinfo
  {author} {\bibfnamefont {Y.}~\bibnamefont {Horibe}}, \bibinfo {author}
  {\bibfnamefont {M.}~\bibnamefont {Tanimura}}, \bibinfo {author}
  {\bibfnamefont {S.}~\bibnamefont {Mori}}, \bibinfo {author} {\bibfnamefont
  {B.}~\bibnamefont {Gao}}, \bibinfo {author} {\bibfnamefont {S.}~\bibnamefont
  {Carr}},\ and\ \bibinfo {author} {\bibfnamefont {S.-W.}\ \bibnamefont
  {Cheong}},\ }\bibfield  {title} {\bibinfo {title} {Direct observation of the
  proliferation of ferroelectric loop domains and vortex-antivortex pairs},\
  }\bibfield  {journal} {\bibinfo  {journal} {Physical Review Letters}\
  }\textbf {\bibinfo {volume} {108}},\ \href
  {https://doi.org/10.1103/physrevlett.108.167603}
  {10.1103/physrevlett.108.167603} (\bibinfo {year} {2012})\BibitemShut
  {NoStop}%
\bibitem [{\citenamefont {Aken}\ \emph {et~al.}(2004)\citenamefont {Aken},
  \citenamefont {Palstra}, \citenamefont {Filippetti},\ and\ \citenamefont
  {Spaldin}}]{Aken2004}%
  \BibitemOpen
  \bibfield  {author} {\bibinfo {author} {\bibfnamefont {B.~B.~V.}\
  \bibnamefont {Aken}}, \bibinfo {author} {\bibfnamefont {T.~T.}\ \bibnamefont
  {Palstra}}, \bibinfo {author} {\bibfnamefont {A.}~\bibnamefont
  {Filippetti}},\ and\ \bibinfo {author} {\bibfnamefont {N.~A.}\ \bibnamefont
  {Spaldin}},\ }\bibfield  {title} {\bibinfo {title} {The origin of
  ferroelectricity in magnetoelectric \ce{YMnO3}},\ }\href
  {https://doi.org/10.1038/nmat1080} {\bibfield  {journal} {\bibinfo  {journal}
  {Nature Materials}\ }\textbf {\bibinfo {volume} {3}},\ \bibinfo {pages} {164}
  (\bibinfo {year} {2004})}\BibitemShut {NoStop}%
\bibitem [{\citenamefont {Choi}\ \emph {et~al.}(2010)\citenamefont {Choi},
  \citenamefont {Horibe}, \citenamefont {Yi}, \citenamefont {Choi},
  \citenamefont {Wu},\ and\ \citenamefont {Cheong}}]{Choi2010}%
  \BibitemOpen
  \bibfield  {author} {\bibinfo {author} {\bibfnamefont {T.}~\bibnamefont
  {Choi}}, \bibinfo {author} {\bibfnamefont {Y.}~\bibnamefont {Horibe}},
  \bibinfo {author} {\bibfnamefont {H.~T.}\ \bibnamefont {Yi}}, \bibinfo
  {author} {\bibfnamefont {Y.~J.}\ \bibnamefont {Choi}}, \bibinfo {author}
  {\bibfnamefont {W.}~\bibnamefont {Wu}},\ and\ \bibinfo {author}
  {\bibfnamefont {S.-W.}\ \bibnamefont {Cheong}},\ }\bibfield  {title}
  {\bibinfo {title} {Insulating interlocked ferroelectric and structural
  antiphase domain walls in multiferroic \ce{YMnO3}},\ }\href
  {https://doi.org/10.1038/nmat2632} {\bibfield  {journal} {\bibinfo  {journal}
  {Nature Materials}\ }\textbf {\bibinfo {volume} {9}},\ \bibinfo {pages} {253}
  (\bibinfo {year} {2010})}\BibitemShut {NoStop}%
\bibitem [{\citenamefont {Wu}\ \emph {et~al.}(2012)\citenamefont {Wu},
  \citenamefont {Horibe}, \citenamefont {Lee}, \citenamefont {Cheong},\ and\
  \citenamefont {Guest}}]{Wu2012}%
  \BibitemOpen
  \bibfield  {author} {\bibinfo {author} {\bibfnamefont {W.}~\bibnamefont
  {Wu}}, \bibinfo {author} {\bibfnamefont {Y.}~\bibnamefont {Horibe}}, \bibinfo
  {author} {\bibfnamefont {N.}~\bibnamefont {Lee}}, \bibinfo {author}
  {\bibfnamefont {S.-W.}\ \bibnamefont {Cheong}},\ and\ \bibinfo {author}
  {\bibfnamefont {J.~R.}\ \bibnamefont {Guest}},\ }\bibfield  {title} {\bibinfo
  {title} {Conduction of topologically protected charged ferroelectric domain
  walls},\ }\bibfield  {journal} {\bibinfo  {journal} {Physical Review
  Letters}\ }\textbf {\bibinfo {volume} {108}},\ \href
  {https://doi.org/10.1103/physrevlett.108.077203}
  {10.1103/physrevlett.108.077203} (\bibinfo {year} {2012})\BibitemShut
  {NoStop}%
\bibitem [{\citenamefont {Roede}\ \emph {et~al.}(2021)\citenamefont {Roede},
  \citenamefont {Mosberg}, \citenamefont {Evans}, \citenamefont {Bourret},
  \citenamefont {Yan}, \citenamefont {van Helvoort},\ and\ \citenamefont
  {Meier}}]{Roede2021}%
  \BibitemOpen
  \bibfield  {author} {\bibinfo {author} {\bibfnamefont {E.~D.}\ \bibnamefont
  {Roede}}, \bibinfo {author} {\bibfnamefont {A.~B.}\ \bibnamefont {Mosberg}},
  \bibinfo {author} {\bibfnamefont {D.~M.}\ \bibnamefont {Evans}}, \bibinfo
  {author} {\bibfnamefont {E.}~\bibnamefont {Bourret}}, \bibinfo {author}
  {\bibfnamefont {Z.}~\bibnamefont {Yan}}, \bibinfo {author} {\bibfnamefont
  {A.~T.~J.}\ \bibnamefont {van Helvoort}},\ and\ \bibinfo {author}
  {\bibfnamefont {D.}~\bibnamefont {Meier}},\ }\bibfield  {title} {\bibinfo
  {title} {Contact-free reversible switching of improper ferroelectric domains
  by electron and ion irradiation},\ }\href {https://doi.org/10.1063/5.0038909}
  {\bibfield  {journal} {\bibinfo  {journal} {{APL} Materials}\ }\textbf
  {\bibinfo {volume} {9}},\ \bibinfo {pages} {021105} (\bibinfo {year}
  {2021})}\BibitemShut {NoStop}%
\bibitem [{\citenamefont {Vasudevan}\ \emph {et~al.}()\citenamefont
  {Vasudevan}, \citenamefont {Morozovska}, \citenamefont {Eliseev},
  \citenamefont {Britson}, \citenamefont {Yang}, \citenamefont {Chu},
  \citenamefont {Maksymovych}, \citenamefont {Chen}, \citenamefont
  {Nagarajan},\ and\ \citenamefont {Kalinin}}]{Vasudevan2012}%
  \BibitemOpen
  \bibfield  {author} {\bibinfo {author} {\bibfnamefont {R.~K.}\ \bibnamefont
  {Vasudevan}}, \bibinfo {author} {\bibfnamefont {A.~N.}\ \bibnamefont
  {Morozovska}}, \bibinfo {author} {\bibfnamefont {E.~A.}\ \bibnamefont
  {Eliseev}}, \bibinfo {author} {\bibfnamefont {J.}~\bibnamefont {Britson}},
  \bibinfo {author} {\bibfnamefont {J.-C.}\ \bibnamefont {Yang}}, \bibinfo
  {author} {\bibfnamefont {Y.-H.}\ \bibnamefont {Chu}}, \bibinfo {author}
  {\bibfnamefont {P.}~\bibnamefont {Maksymovych}}, \bibinfo {author}
  {\bibfnamefont {L.~Q.}\ \bibnamefont {Chen}}, \bibinfo {author}
  {\bibfnamefont {V.}~\bibnamefont {Nagarajan}},\ and\ \bibinfo {author}
  {\bibfnamefont {S.~V.}\ \bibnamefont {Kalinin}},\ }\bibfield  {title}
  {\bibinfo {title} {Domain wall geometry controls conduction in
  ferroelectrics},\ }\href {https://doi.org/10.1021/nl302382k} {\bibfield
  {journal} {\bibinfo  {journal} {Nano Letters}\ }\textbf {\bibinfo {volume}
  {12}},\ \bibinfo {pages} {5524}}\BibitemShut {NoStop}%
\bibitem [{\citenamefont {Abdollahi}\ \emph {et~al.}(2019)\citenamefont
  {Abdollahi}, \citenamefont {Domingo}, \citenamefont {Arias},\ and\
  \citenamefont {Catalan}}]{Abdollahi2019}%
  \BibitemOpen
  \bibfield  {author} {\bibinfo {author} {\bibfnamefont {A.}~\bibnamefont
  {Abdollahi}}, \bibinfo {author} {\bibfnamefont {N.}~\bibnamefont {Domingo}},
  \bibinfo {author} {\bibfnamefont {I.}~\bibnamefont {Arias}},\ and\ \bibinfo
  {author} {\bibfnamefont {G.}~\bibnamefont {Catalan}},\ }\bibfield  {title}
  {\bibinfo {title} {Converse flexoelectricity yields large piezoresponse force
  microscopy signals in non-piezoelectric materials},\ }\bibfield  {journal}
  {\bibinfo  {journal} {Nature Communications}\ }\textbf {\bibinfo {volume}
  {10}},\ \href {https://doi.org/10.1038/s41467-019-09266-y}
  {10.1038/s41467-019-09266-y} (\bibinfo {year} {2019})\BibitemShut {NoStop}%
\bibitem [{\citenamefont {Gregg}(2022)}]{Gregg2022}%
  \BibitemOpen
  \bibfield  {author} {\bibinfo {author} {\bibfnamefont {J.~M.}\ \bibnamefont
  {Gregg}},\ }\bibfield  {title} {\bibinfo {title} {A perspective on conducting
  domain walls and possibilities for ephemeral electronics},\ }\href
  {https://doi.org/10.1063/5.0079738} {\bibfield  {journal} {\bibinfo
  {journal} {Applied Physics Letters}\ }\textbf {\bibinfo {volume} {120}},\
  \bibinfo {pages} {010501} (\bibinfo {year} {2022})}\BibitemShut {NoStop}%
\bibitem [{\citenamefont {Schaab}\ \emph {et~al.}(2015)\citenamefont {Schaab},
  \citenamefont {Cano}, \citenamefont {Lilienblum}, \citenamefont {Yan},
  \citenamefont {Bourret}, \citenamefont {Ramesh}, \citenamefont {Fiebig},\
  and\ \citenamefont {Meier}}]{Schaab2015}%
  \BibitemOpen
  \bibfield  {author} {\bibinfo {author} {\bibfnamefont {J.}~\bibnamefont
  {Schaab}}, \bibinfo {author} {\bibfnamefont {A.}~\bibnamefont {Cano}},
  \bibinfo {author} {\bibfnamefont {M.}~\bibnamefont {Lilienblum}}, \bibinfo
  {author} {\bibfnamefont {Z.}~\bibnamefont {Yan}}, \bibinfo {author}
  {\bibfnamefont {E.}~\bibnamefont {Bourret}}, \bibinfo {author} {\bibfnamefont
  {R.}~\bibnamefont {Ramesh}}, \bibinfo {author} {\bibfnamefont
  {M.}~\bibnamefont {Fiebig}},\ and\ \bibinfo {author} {\bibfnamefont
  {D.}~\bibnamefont {Meier}},\ }\bibfield  {title} {\bibinfo {title}
  {Optimization of electronic domain-wall properties by aliovalent cation
  substitution},\ }\href {https://doi.org/10.1002/aelm.201500195} {\bibfield
  {journal} {\bibinfo  {journal} {Advanced Electronic Materials}\ }\textbf
  {\bibinfo {volume} {2}},\ \bibinfo {pages} {1500195} (\bibinfo {year}
  {2015})}\BibitemShut {NoStop}%
\bibitem [{\citenamefont {Schulthei{\ss}}\ \emph {et~al.}(2021)\citenamefont
  {Schulthei{\ss}}, \citenamefont {Rojac},\ and\ \citenamefont
  {Meier}}]{Schultheis2021}%
  \BibitemOpen
  \bibfield  {author} {\bibinfo {author} {\bibfnamefont {J.}~\bibnamefont
  {Schulthei{\ss}}}, \bibinfo {author} {\bibfnamefont {T.}~\bibnamefont
  {Rojac}},\ and\ \bibinfo {author} {\bibfnamefont {D.}~\bibnamefont {Meier}},\
  }\bibfield  {title} {\bibinfo {title} {Unveiling alternating current
  electronic properties at ferroelectric domain walls},\ }\href
  {https://doi.org/10.1002/aelm.202100996} {\bibfield  {journal} {\bibinfo
  {journal} {Advanced Electronic Materials}\ ,\ \bibinfo {pages} {2100996}}
  (\bibinfo {year} {2021})}\BibitemShut {NoStop}%
\bibitem [{\citenamefont {Llewellyn-Jones}(1957)}]{LJ1957}%
  \BibitemOpen
  \bibfield  {author} {\bibinfo {author} {\bibfnamefont {F.}~\bibnamefont
  {Llewellyn-Jones}},\ }\href@noop {} {\emph {\bibinfo {title} {The Physics of
  Electrical Contacts}}}\ (\bibinfo  {publisher} {Clarendon Press},\ \bibinfo
  {year} {1957})\BibitemShut {NoStop}%
\bibitem [{\citenamefont {Holm}(1958)}]{holm1958}%
  \BibitemOpen
  \bibfield  {author} {\bibinfo {author} {\bibfnamefont {R.}~\bibnamefont
  {Holm}},\ }\href@noop {} {\emph {\bibinfo {title} {Electric Contacts
  Handbook}}}\ (\bibinfo  {publisher} {Springer},\ \bibinfo {year}
  {1958})\BibitemShut {NoStop}%
\bibitem [{\citenamefont {Werner}\ \emph {et~al.}(2017)\citenamefont {Werner},
  \citenamefont {Herr}, \citenamefont {Buse}, \citenamefont {Sturman},
  \citenamefont {Soergel}, \citenamefont {Razzaghi},\ and\ \citenamefont
  {Breunig}}]{Werner2017}%
  \BibitemOpen
  \bibfield  {author} {\bibinfo {author} {\bibfnamefont {C.~S.}\ \bibnamefont
  {Werner}}, \bibinfo {author} {\bibfnamefont {S.~J.}\ \bibnamefont {Herr}},
  \bibinfo {author} {\bibfnamefont {K.}~\bibnamefont {Buse}}, \bibinfo {author}
  {\bibfnamefont {B.}~\bibnamefont {Sturman}}, \bibinfo {author} {\bibfnamefont
  {E.}~\bibnamefont {Soergel}}, \bibinfo {author} {\bibfnamefont
  {C.}~\bibnamefont {Razzaghi}},\ and\ \bibinfo {author} {\bibfnamefont
  {I.}~\bibnamefont {Breunig}},\ }\bibfield  {title} {\bibinfo {title} {Large
  and accessible conductivity of charged domain walls in lithium niobate},\
  }\bibfield  {journal} {\bibinfo  {journal} {Scientific Reports}\ }\textbf
  {\bibinfo {volume} {7}},\ \href {https://doi.org/10.1038/s41598-017-09703-2}
  {10.1038/s41598-017-09703-2} (\bibinfo {year} {2017})\BibitemShut {NoStop}%
\bibitem [{\citenamefont {Song}\ \emph
  {et~al.}(2021{\natexlab{a}})\citenamefont {Song}, \citenamefont {Zhou},\ and\
  \citenamefont {Huey}}]{Song2021}%
  \BibitemOpen
  \bibfield  {author} {\bibinfo {author} {\bibfnamefont {J.}~\bibnamefont
  {Song}}, \bibinfo {author} {\bibfnamefont {Y.}~\bibnamefont {Zhou}},\ and\
  \bibinfo {author} {\bibfnamefont {B.~D.}\ \bibnamefont {Huey}},\ }\bibfield
  {title} {\bibinfo {title} {3d structure{\textendash}property correlations of
  electronic and energy materials by tomographic atomic force microscopy},\
  }\href {https://doi.org/10.1063/5.0040984} {\bibfield  {journal} {\bibinfo
  {journal} {Applied Physics Letters}\ }\textbf {\bibinfo {volume} {118}},\
  \bibinfo {pages} {080501} (\bibinfo {year} {2021}{\natexlab{a}})}\BibitemShut
  {NoStop}%
\bibitem [{\citenamefont {Song}\ \emph
  {et~al.}(2021{\natexlab{b}})\citenamefont {Song}, \citenamefont {Zhuang},
  \citenamefont {Martin}, \citenamefont {Ortiz-Flores}, \citenamefont {Paudel},
  \citenamefont {Yarotski}, \citenamefont {Hu}, \citenamefont {Chen},\ and\
  \citenamefont {Huey}}]{Song2021a}%
  \BibitemOpen
  \bibfield  {author} {\bibinfo {author} {\bibfnamefont {J.}~\bibnamefont
  {Song}}, \bibinfo {author} {\bibfnamefont {S.}~\bibnamefont {Zhuang}},
  \bibinfo {author} {\bibfnamefont {M.}~\bibnamefont {Martin}}, \bibinfo
  {author} {\bibfnamefont {L.~A.}\ \bibnamefont {Ortiz-Flores}}, \bibinfo
  {author} {\bibfnamefont {B.}~\bibnamefont {Paudel}}, \bibinfo {author}
  {\bibfnamefont {D.}~\bibnamefont {Yarotski}}, \bibinfo {author}
  {\bibfnamefont {J.}~\bibnamefont {Hu}}, \bibinfo {author} {\bibfnamefont
  {A.}~\bibnamefont {Chen}},\ and\ \bibinfo {author} {\bibfnamefont {B.~D.}\
  \bibnamefont {Huey}},\ }\bibfield  {title} {\bibinfo {title}
  {Interfacial-strain-controlled ferroelectricity in self-assembled \ce{BiFeO3}
  nanostructures},\ }\href {https://doi.org/10.1002/adfm.202102311} {\bibfield
  {journal} {\bibinfo  {journal} {Advanced Functional Materials}\ }\textbf
  {\bibinfo {volume} {31}},\ \bibinfo {pages} {2102311} (\bibinfo {year}
  {2021}{\natexlab{b}})}\BibitemShut {NoStop}%
\bibitem [{\citenamefont {Steffes}\ \emph {et~al.}(2019)\citenamefont
  {Steffes}, \citenamefont {Ristau}, \citenamefont {Ramesh},\ and\
  \citenamefont {Huey}}]{Steffes2019}%
  \BibitemOpen
  \bibfield  {author} {\bibinfo {author} {\bibfnamefont {J.~J.}\ \bibnamefont
  {Steffes}}, \bibinfo {author} {\bibfnamefont {R.~A.}\ \bibnamefont {Ristau}},
  \bibinfo {author} {\bibfnamefont {R.}~\bibnamefont {Ramesh}},\ and\ \bibinfo
  {author} {\bibfnamefont {B.~D.}\ \bibnamefont {Huey}},\ }\bibfield  {title}
  {\bibinfo {title} {Thickness scaling of ferroelectricity in \ce{BiFeO3} by
  tomographic atomic force microscopy},\ }\href
  {https://doi.org/10.1073/pnas.1806074116} {\bibfield  {journal} {\bibinfo
  {journal} {Proceedings of the National Academy of Sciences}\ }\textbf
  {\bibinfo {volume} {116}},\ \bibinfo {pages} {2413} (\bibinfo {year}
  {2019})}\BibitemShut {NoStop}%
\bibitem [{\citenamefont {Rueden}\ \emph {et~al.}(2017)\citenamefont {Rueden},
  \citenamefont {Schindelin}, \citenamefont {Hiner}, \citenamefont {DeZonia},
  \citenamefont {Walter}, \citenamefont {Arena},\ and\ \citenamefont
  {Eliceiri}}]{Rueden2017}%
  \BibitemOpen
  \bibfield  {author} {\bibinfo {author} {\bibfnamefont {C.~T.}\ \bibnamefont
  {Rueden}}, \bibinfo {author} {\bibfnamefont {J.}~\bibnamefont {Schindelin}},
  \bibinfo {author} {\bibfnamefont {M.~C.}\ \bibnamefont {Hiner}}, \bibinfo
  {author} {\bibfnamefont {B.~E.}\ \bibnamefont {DeZonia}}, \bibinfo {author}
  {\bibfnamefont {A.~E.}\ \bibnamefont {Walter}}, \bibinfo {author}
  {\bibfnamefont {E.~T.}\ \bibnamefont {Arena}},\ and\ \bibinfo {author}
  {\bibfnamefont {K.~W.}\ \bibnamefont {Eliceiri}},\ }\bibfield  {title}
  {\bibinfo {title} {{ImageJ}2: {ImageJ} for the next generation of scientific
  image data},\ }\bibfield  {journal} {\bibinfo  {journal} {{BMC}
  Bioinformatics}\ }\textbf {\bibinfo {volume} {18}},\ \href
  {https://doi.org/10.1186/s12859-017-1934-z} {10.1186/s12859-017-1934-z}
  (\bibinfo {year} {2017})\BibitemShut {NoStop}%
\bibitem [{\citenamefont {Schindelin}\ \emph {et~al.}(2012)\citenamefont
  {Schindelin}, \citenamefont {Arganda-Carreras}, \citenamefont {Frise},
  \citenamefont {Kaynig}, \citenamefont {Longair}, \citenamefont {Pietzsch},
  \citenamefont {Preibisch}, \citenamefont {Rueden}, \citenamefont {Saalfeld},
  \citenamefont {Schmid}, \citenamefont {Tinevez}, \citenamefont {White},
  \citenamefont {Hartenstein}, \citenamefont {Eliceiri}, \citenamefont
  {Tomancak},\ and\ \citenamefont {Cardona}}]{Schindelin2012}%
  \BibitemOpen
  \bibfield  {author} {\bibinfo {author} {\bibfnamefont {J.}~\bibnamefont
  {Schindelin}}, \bibinfo {author} {\bibfnamefont {I.}~\bibnamefont
  {Arganda-Carreras}}, \bibinfo {author} {\bibfnamefont {E.}~\bibnamefont
  {Frise}}, \bibinfo {author} {\bibfnamefont {V.}~\bibnamefont {Kaynig}},
  \bibinfo {author} {\bibfnamefont {M.}~\bibnamefont {Longair}}, \bibinfo
  {author} {\bibfnamefont {T.}~\bibnamefont {Pietzsch}}, \bibinfo {author}
  {\bibfnamefont {S.}~\bibnamefont {Preibisch}}, \bibinfo {author}
  {\bibfnamefont {C.}~\bibnamefont {Rueden}}, \bibinfo {author} {\bibfnamefont
  {S.}~\bibnamefont {Saalfeld}}, \bibinfo {author} {\bibfnamefont
  {B.}~\bibnamefont {Schmid}}, \bibinfo {author} {\bibfnamefont {J.-Y.}\
  \bibnamefont {Tinevez}}, \bibinfo {author} {\bibfnamefont {D.~J.}\
  \bibnamefont {White}}, \bibinfo {author} {\bibfnamefont {V.}~\bibnamefont
  {Hartenstein}}, \bibinfo {author} {\bibfnamefont {K.}~\bibnamefont
  {Eliceiri}}, \bibinfo {author} {\bibfnamefont {P.}~\bibnamefont {Tomancak}},\
  and\ \bibinfo {author} {\bibfnamefont {A.}~\bibnamefont {Cardona}},\
  }\bibfield  {title} {\bibinfo {title} {Fiji: an open-source platform for
  biological-image analysis},\ }\href {https://doi.org/10.1038/nmeth.2019}
  {\bibfield  {journal} {\bibinfo  {journal} {Nature Methods}\ }\textbf
  {\bibinfo {volume} {9}},\ \bibinfo {pages} {676} (\bibinfo {year}
  {2012})}\BibitemShut {NoStop}%
\bibitem [{\citenamefont {Ayachit}(2015)}]{Ayachit2015}%
  \BibitemOpen
  \bibfield  {author} {\bibinfo {author} {\bibfnamefont {U.}~\bibnamefont
  {Ayachit}},\ }\href@noop {} {\emph {\bibinfo {title} {The ParaView guide :
  updated for ParaView version 4.3}}}\ (\bibinfo  {publisher} {Kitware},\
  \bibinfo {address} {Clifton Park, New York},\ \bibinfo {year}
  {2015})\BibitemShut {NoStop}%
\end{thebibliography}%

\renewcommand{\thefigure}{S\arabic{figure}}

\setcounter{figure}{0}

\section{Supplementary Information}%

\subsubsection{Note 1: Details of the finite-element modeling.}

The current flow in presence of the extracted domain wall structure (Fig. 3a) is simulated in two steps. First we obtain the distribution of the polarization $P(x,y,z)$ within the domain wall network of \ce{ErMnO3} based on the extracted 3D domain wall network data. For this, we minimize the Landau energy for a uniaxial ferroelectric, $$\alpha P^2/2 + \beta P^4/4 +\delta\left[ (\partial P / \partial x)^2 + (\partial P / \partial y)^2  + (\partial P / \partial z)^2\right]/2,$$ by solving the Euler-Lagrange equation for the polarization, $$\alpha P + \beta P^3 -\delta \left[ \partial^2 P / \partial x^2 +  \partial^2 P / \partial y^2 + \partial^2 P / \partial z^2 \right] = 0$$ (correlation tensor $\delta$ is assumed to be isotropic). We use dimensionless units for polarization by setting $\alpha = -1, \beta =1, \delta = \xi^2 /2$, where $\xi=\SI{1}{\nano\metre}$ is the domain wall thickness, which corresponds to the polarization profile at straight domain walls $P(x)=\tanh{x/\xi}$. We start our simulations of the polarization in the domain wall network from the initial distribution $P(x,y,z)=\pm 1$, with the polarization having opposite signs in any two neighbouring domains (as defined by the extracted domain wall network), and use iterative finite-elements methods implemented in COMSOL Multiphysics with imposed boundary condition $P(x,y,z)=0$ at the domain walls to solve the Euler-Lagrange equations for polarization. The resulting distribution of polarization $P(x,y,z)$ is then fed into the condition for the non-divergence of the electric current, $$\nabla \cdot \vec{j} = -\nabla (\sigma \nabla \phi) = 0,$$  where $\phi$ is the electric potential and $\sigma = \exp{(\partial P /\partial x)/s}$ is the local conductivity normalized to the bulk conductivity. We use $s=\xi/5$ in our simulations, which results in the maximal possible value of conductivity in the centre of the straight tail-to-tail domain wall $\sigma_\mathrm{DW}^\mathrm{max}=exp(5)\approx150$. The equation is solved with respect to the electric potential $\phi(x,y,z)$ with imposed boundary conditions $\phi=1$ at the AFM tip-surface contact (represented by a disc of radius $r=\SI{4}{\nano\metre}$) and $\phi=0$ at the bottom electrode. The dimensions of the system used in the simulations are $750$×$500$×$\SI{340}{\nano\metre}$ respectively in $x$, $y$ and $z$ directions shown in Fig. 3a; these dimensions are scaled down 10 times with respect to the ones extracted from the real system in order to accommodate for the limited available computational power. The tetrahedral finite-element mesh with quadratic discretization is used, with adaptive finite element size ranging from \SI{2}{\nano\metre} at the AFM tip-surface contact to \SI{10}{\nano\metre}  at the domain walls and to \SI{50}{\nano\metre}  in the bulk domains.

For quantitative calculations of conductance at straight and curved domain walls summarized in Figs. 3b and 4d, we use the analytic solution for the polarization at the wall:
$$ P (\tilde{x}) = \tanh{(\tilde{x}/\tilde{\xi})},$$
where $\tilde{x}$ is the direction perpendicular to the wall and $\xi=\SI{1}{\nano\metre}$ is the domain wall thickness – radius of the domain wall curvature much larger than $\xi$ is implied. The analytic solution is fed into the condition for the non-divergence of the electric current, $\nabla \cdot \vec{j} = -\nabla (\sigma \nabla \phi) = 0$ where $\phi$ is the electric potential, $\sigma = \exp{ ([\partial P/\partial x]/s) H(d-z) }$
) is the local conductivity normalized to the bulk conductivity, $H$ is the Heaviside function and $d$ is the domain wall conductivity cut-off distance from the surface (the cut-off distance analysis is only made for straight walls, Fig. 3b). The equation is solved with respect to the electric potential $\phi(x,y,z)$ with imposed boundary conditions $\phi=1$ at the AFM tip-surface contact (represented by a disc of radius $r=\SI{4}{\nano\metre}$) and $\phi=0$ at the bottom electrode. The total current flow is measured by integrating $\vec{j}=-\sigma \nabla \phi$ over the bottom electrode; the relative conductance $G/G_\mathrm{bulk}$ is calculated by dividing the obtained current flow over the value calculated in the absence of the domain wall. The dimensions of the system used in the simulations are $500$×$250$×$\SI{200}{\nano\metre}$ respectively in $x$, $y$ and $z$ directions; the used finite-element mesh is tetrahedral with quadratic discretization, with size ranging from \SI{2}{\nano\metre} at the AFM tip-surface contact, domain wall and bottom electrode to \SI{20}{\nano\metre}  in the bulk domains.

\begin{figure}
\includegraphics[width=0.48\textwidth]{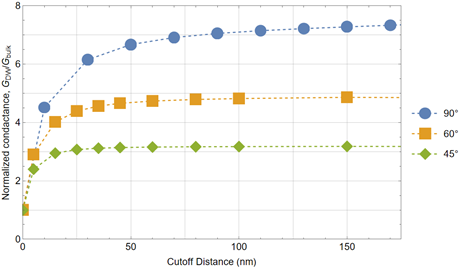}
\caption{\label{fig:S2} Simulated conductance through a conductive domain wall (relative to bulk conductivity) for artificial domain wall cut off at varying depth, for the straight domain walls inclined by \SI{90}{\degree}, \SI{60}{\degree} and \SI{45}{\degree} with respect to the surface. }
\end{figure}

\begin{figure*}
\includegraphics[width=\textwidth]{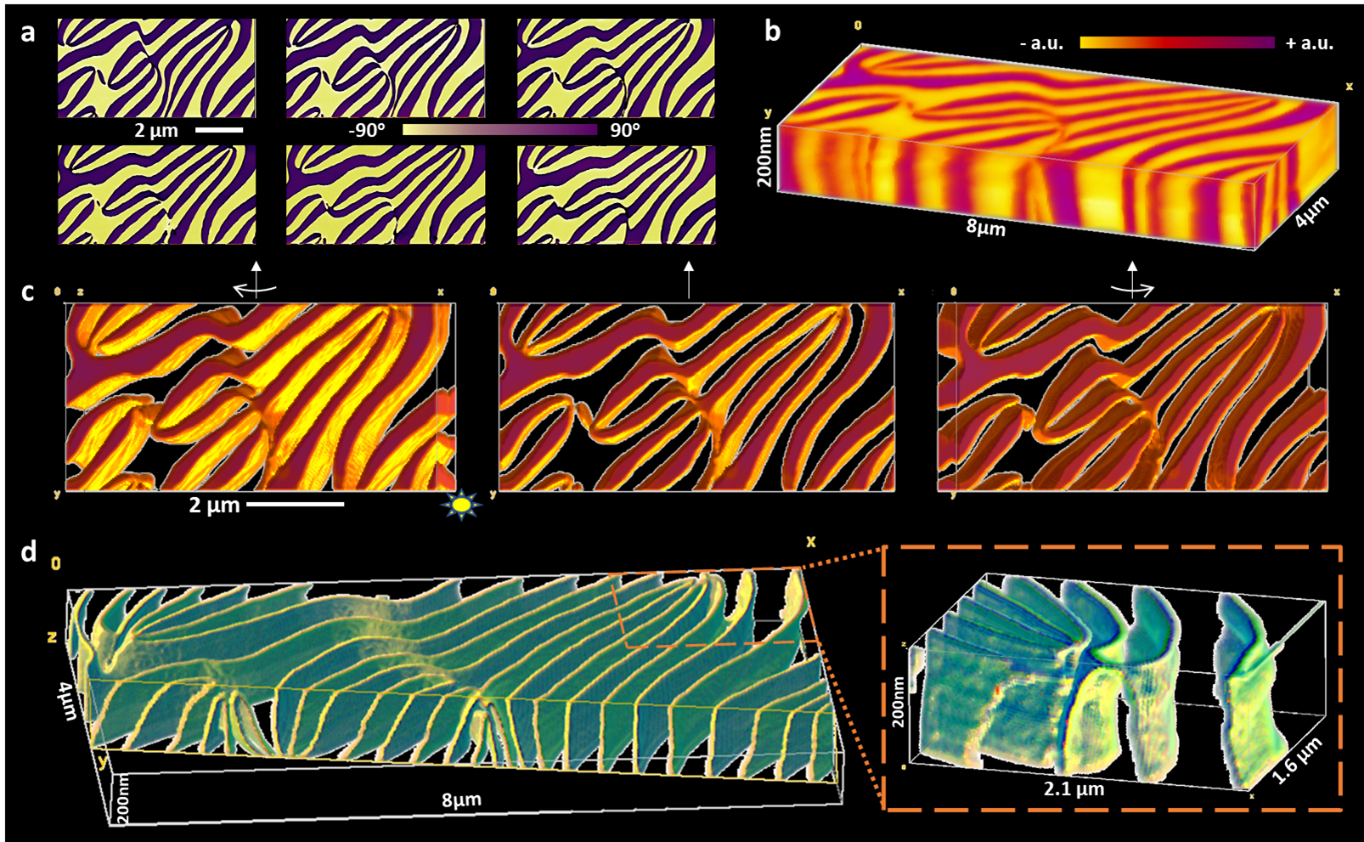}
\caption{\label{fig:S1} Tomographic AFM reveals 3D curvature of domain patterns in \ce{ErMnO3}. Representative lateral PFM phase images acquired during Tomographic AFM (a), along with equivalent lateral PFM amplitude signals, enables reconstruction of the piezoresponse throughout an excavated volume (b). Digitally visualizing only the positive domains, viewed from various perspectives and with oblique illumination, reveals extensive sub-surface variations in the domain structure (c). The local slopes and curvature are especially apparent with perspective views and cross-sections of only the domain walls (d, and magnified inset).}
\end{figure*}
\end{document}